\pdfoutput=1

\documentclass[journal]{IEEEtran}

\ifCLASSINFOpdf
\else
\fi

\usepackage{url}

\usepackage{epsf,epsfig}
\usepackage{warpcol}
\usepackage{multirow}
\usepackage{tikz}
\usetikzlibrary{calc, positioning, arrows, backgrounds}
\usepackage{pgf}
\newcommand{\mycaption}[1]{\caption{#1}}
\newcommand{\drawU}[1]{
		\draw[thick] (#1.north east) -- (#1.north west) --
			  (#1.south west) -- (#1.south east);
}

\newcommand{\legendforblocksV}[0]{%
\begin{tikzpicture}[legendbox/.style={inner sep=1.5ex,draw=black,thick}]
	\begin{scope}[node distance=4ex]
		\node at (1.5cm, 1cm) {Legend};
		\node[] (leg0) at (0,0) {};
		\node[below of=leg0, inner sep=1.5ex,st_reducible,alias=leg1] 
			 (reduciblebox) {};
		\drawU{reduciblebox}
		\node[right of=reduciblebox, right] {reducible block};

		\node[below of=leg1, inner sep=1.5ex,st_irreducible,alias=leg2] (irreduciblebox) {};
		\drawU{irreduciblebox}
		\node[right of=irreduciblebox, right] (irrlabel) 
		     {irreducible block};
		\node[below of=leg2, inner sep=1.5ex,st_singleton,alias=leg3] (singletonbox) {};
		\drawU{singletonbox}
		\node[right of=singletonbox, right] {singleton block};

		\node[below of=leg3, st_block_fwd_id_label, alias=leg4, yshift=-1.3cm] (block_label) {x}; 
		\node[st_block_bwd_id_label,node distance=0ex] (block_bwd_id_label) at (block_label.north east) {y};
		\node[right of=block_label, right, text width=4cm]{label of block $x$ with $\bwdid{x}=y$};

		\node[below of=block_label, inner sep=1.5ex,st_interval,alias=leg5, yshift=-0.2cm] 
			 (omega interval box) {};
		\drawU{omega interval box}
		\node[right of=omega interval box, right] {$\omega$-interval};

	\end{scope}
\end{tikzpicture}%
}

\newcommand{\getdeltax}[1]{
	\coordinate (deltax) at ($#1*(text0.west)-#1*(text0.east)+#1*2*(\pgflinewidth,0)$);
}

\newcommand{\markintervalleft}[5]{%
\begin{tikzpicture}[overlay, remember picture,font=\tt,show background rectangle]
	\getdeltax{#3}
	\draw[#5] (#4#1.north west) ++ ($(0,0)-(deltax)$) -- ++ (deltax) 
	           -- (#4#2.south west) -- ++ ($(0,0)-(deltax)$);
\end{tikzpicture}%
}

\newcommand{\markintervalright}[5]{%
\begin{tikzpicture}[overlay,remember picture, font=\tt, st_mark_interval_right]
	\getdeltax{#3}
	\draw[#5] (#4#1.north east) -- ++ (deltax) -- 
	          ($(#4#2.south east)+ (deltax)$) -- 
			  ++ ($(0,0)-(deltax)$);
	\draw (#4#1.north east) -- ++ (deltax) -- 
	          ($(#4#2.south east)+ (deltax)$) -- 
			  ++ ($(0,0)-(deltax)$);
\end{tikzpicture}%
}

\newcommand{\markfringe}[4]{%
\begin{tikzpicture}[overlay,remember picture, font=\tt]
	\getdeltax{1}
	\draw[#4] ($(text#1.north west)-#2*(deltax)-#2*(\pgflinewidth,0)$) 
			  rectangle ($(text#1.south west)-#2*(deltax)-#3*(deltax)-#2*(\pgflinewidth,0)$);
\end{tikzpicture}%
}

\newcommand{\markfixedfringes}[4]{
	\foreach \idx [count=\i from 0] in #1{
		\pgfmathparse{array(#2,\i)}\let\fringedepth\pgfmathresult
		\markfringe{\idx}{\fringedepth}{#3}{#4}	
	}
}

\newcommand{\markvarfringes}[4]{
	\foreach \idx [count=\i from 0] in #1{
		\pgfmathparse{array(#2,\i)}\let\fringedepth\pgfmathresult
		\pgfmathparse{array(#3,\i)}\let\fringelen\pgfmathresult
		\markfringe{\idx}{\fringedepth}{\fringelen}{#4}	
	}
}

\newcommand{\intervalanchor}[4]{
	\tikz[overlay,remember picture] \coordinate (#4) at ($(#3#1.east)!0.5!(#3#2.east)$);
	\tikz[overlay,remember picture] \coordinate (#4_start) at (#3#1.north east);
	\tikz[overlay,remember picture] \coordinate (#4_start_east) at (#3#1.north east);
	\tikz[overlay,remember picture] \coordinate (#4_end) at (#3#2.south east);
}

\newcommand{\intervaledge}[3]{%
	\begin{tikzpicture}[overlay, remember picture, st_block_info]
		\coordinate (block1_coord) at ($(block#1)-(3.5ex,0.5ex)$);
		\coordinate (block2_coord) at ($(block#2)-(3.5ex,-0.5ex)$);
		\ifnum #1 > #2
			\draw[st_interval_edge, #3] (block1_coord) to[bend left=30] (block2_coord);
		\else
			\ifnum #1 < #2
				\draw[st_interval_edge, #3] (block1_coord) to[bend right=30] (block2_coord);
			\else %
				\draw[fill=black] ($(block1_coord)-(0.5ex,-0.5ex)$) circle (2pt);
			\fi	
		\fi
	\end{tikzpicture}%
}
\newcommand{\intervaledgereducible}[3]{%
	\begin{tikzpicture}[overlay, remember picture, st_block_info]
		\coordinate (len#1) at ($(block#1_end)-(block#1_start)$);
		\coordinate (line_height) at ($(i0.north)-(i1.north)$);
		\coordinate (block1_coord2) at ($(block#1)-(3.5ex,0ex)$);
		\coordinate (block2_coord2) at ($(block#2_start)-(3.5ex,0ex)-#3*(line_height)+1/2.0*(len#1)$);
		\ifnum #1 > #2
			\draw[st_reducible_edge] (block1_coord2) to[bend left=30] (block2_coord2);
		\else
			\ifnum #1 < #2
				\draw[st_reducible_edge] (block1_coord2) to[bend right=30] (block2_coord2);
			\else %
				\draw[fill=black] ($(block1_coord2)-(0.5ex,-0.5ex)$) circle (2pt);
			\fi	
		\fi	
	\end{tikzpicture}%
}
\newcommand{\blockinfo}[5]{%
\begin{tikzpicture}[overlay, remember picture, st_block_info]
	\ifnum #1 = #4
	\else
		\node[below right,scale=0.7, align=left] (block_info#1) at ($(block#1_start_east)$) 
		 {$\deltax=#2$\ \\$\deltad=#3$};
	\fi
	\node[right, st_block_fwd_id_label,xshift=10ex] (block_fwd_id#1) 
	      at ($(block#1_start_east)$) {#1};
	\draw[st_line_to_block_fwd_id_label] (block#1_start_east) -- (block_fwd_id#1);
	\node[st_block_bwd_id_label] (bwd_block_fwd_id#1) at (block_fwd_id#1.north east) {#5};
\end{tikzpicture}%
}
\newcommand{\externalblockheader}[2]{%
\begin{tikzpicture}[overlay, remember picture, st_block_info,node distance=0ex, inner sep=0.2mm]
	\foreach \item [count=\i from 1, count=\j from 0] in #2{
		\pgfmathparse{#2[\j][0]}\let\ID\pgfmathresult
		\pgfmathparse{#2[\j][1]}\let\DELTAX\pgfmathresult
		\pgfmathparse{#2[\j][2]}\let\DELTAD\pgfmathresult
		\ifnum \j > 0
			\node[below left, st_block_header] 
			     (block_header_#1_\i) at 
				 (block_header_#1_\j.south east)
				 {$(\ID,\DELTAX,\DELTAD)$};
		\else
			\node[below left, st_block_header] (block_header_#1_\i) at (block_fwd_id#1.south west) 
			     {$(\ID,\DELTAX,\DELTAD)$};

		\fi	
	}
\end{tikzpicture}%
}

\newcommand{\drawheaderfwd}[0]{%
\begin{tikzpicture}[remember picture, overlay, every node/.style={inner sep=0cm}]
\coordinate[st_fwd_index_header] (yHeader) at (y0);%
\node[st_elem_i] at (i |- yHeader) {$i$};%
\node[st_elem_sa] at (sa |- yHeader) {$\SUF[i]$};%
\node[st_elem_lcp] at (lcp |- yHeader) {$\LCP[i]$};%
\node[st_elem_bwt] at (bwt |- yHeader) {$\BWT[i]$};%
\coordinate (text_label_pos) at (text0.west |- yHeader);%
\node[st_elem_text] at (text_label_pos) {$\TEXT[\SUF[i]..n]$};%
\end{tikzpicture}%
}
\newcommand{\drawheaderbwd}[0]{%
\begin{tikzpicture}[remember picture, overlay, every node/.style={inner sep=0cm}]
	\coordinate[st_bwd_index_header] (yHeader) at (y0);
	\node[st_elem_i] at (i |- yHeader) {$i$};
	\node[st_elem_bwt] at (bwt |- yHeader) {$\BWT^{\TEXT^{r}}[i]$};
	\node[st_elem_bwdbl] at (bwdbl |- yHeader) {$\bwdbl[i]$};
	\node[st_elem_bwdbf] at (bwdbf |- yHeader) {$\bwdbf[i]$};
	\coordinate (text_label_pos) at (text0.west |- yHeader);
	\node[st_elem_text] at (text_label_pos) {$\TEXT^{r}[\SUF[i]..n]$};
\end{tikzpicture}%
}
\newcommand{\markFwdBlocksInBwdIndex}[0]{
	\foreach \block [count=\bwdId from 0] in \fwdBlocksInBwd{
		\pgfmathparse{int(array(\block,0))}\let\bwdlb\pgfmathresult
		\pgfmathparse{int(array(\block,1)-1)}\let\bwdrb\pgfmathresult
		\pgfmathparse{int(array(\block,2))}\let\fwddepth\pgfmathresult
		\pgfmathparse{int(array(\block,3))}\let\fwdblockid\pgfmathresult
		\markintervalleft{\bwdlb}{\bwdrb}{\fwddepth}{text}{st_interval}%
		\begin{tikzpicture}[overlay, remember picture]%
			\node[st_block_fwd_id_label,opacity=0] (tmp) at (0,0) {\fwdblockid};
			\coordinate (dim) at ($(tmp.east)-(tmp.west)$);
			\coordinate (fwd_block_fwd_id_\bwdlb_\fwddepth_pos) at %
			  			($(text\bwdlb.north east)+\fwddepth*(dim)$);
			\draw[st_line_to_block_fwd_id_label] %
			     (text\bwdlb.north west) -- %
				 (fwd_block_fwd_id_\bwdlb_\fwddepth_pos);
			\node[st_block_fwd_id_label, below] (fwd_block_fwd_id_\bwdlb_\fwddepth) %
			  at (fwd_block_fwd_id_\bwdlb_\fwddepth_pos) {\fwdblockid};
			\node[shape=circle,draw=black,inner sep=0.2mm,fill=black] at 
				(fwd_block_fwd_id_\bwdlb_\fwddepth_pos) {};

			\node[st_block_bwd_id_label] at (fwd_block_fwd_id_\bwdlb_\fwddepth_pos) {\bwdId};
		\end{tikzpicture}%
	}
}
\newcommand{\bwdIndexLayout}[0]{
\tikzstyle{st_i} = [xshift=-0.4cm]\tikzstyle{st_elem_i} = [left]
\tikzstyle{st_lcp} = [xshift=2.3cm]\tikzstyle{st_elem_lcp} = [left,opacity=0]
\tikzstyle{st_bwt} = [xshift=1cm] \tikzstyle{st_elem_bwt} = [left]
\tikzstyle{st_bwdbl} = [xshift=2cm] \tikzstyle{st_elem_bwdbl} = [left]
\tikzstyle{st_bwdbf} = [xshift=3cm] \tikzstyle{st_elem_bwdbf} = [left]
\tikzstyle{st_text} = [xshift=3.4cm] \tikzstyle{st_elem_text} = [right]
}
\newcommand{\printBlAndCBWT}[0]{%
\begin{tikzpicture}[remember picture]
	\begin{scope}[node distance=1.5ex,font=\tt]
		\coordinate (bwdBl0) at (0,0);
		\coordinate (bwdBWT0) at (0,-3ex);
		\node[left=-1.3ex of bwdBl0] {$\bwdbl =\ $};
		\node[left=-1.3ex of bwdBWT0] {$\cL =\ $};
		\foreach \x [count=\i from 0, count=\j from 1] in \bwdBlArray{
			\node[right, right of=bwdBl\i, xshift=-0.2ex] (bwdBl\j) {\x};
			\ifnum \x = 1
				\pgfmathparse{{\bwdBWTArray}[\i]}\let\bwt\pgfmathresult
				\node[text height=2ex, text depth=1ex] at (bwdBl\j |- bwdBWT0) {\bwt};
			\fi	
		}
	\end{scope}
\end{tikzpicture}%
}
\newcounter{minPosCnt}
\newcommand{\printBmAndMinDepth}[0]{%
\begin{tikzpicture}[remember picture]
	\begin{scope}[node distance=1.5ex,font=\tt]
	\setcounter{minPosCnt}{0}
		\coordinate (bwdbm0) at (0,0);
		\coordinate (minDepth0) at (0,-3ex);
		\coordinate (XY) at (bwdbm0 |- minDepth0);
		\node[left=-1.3ex of bwdbm0] {$\bwdbm =\ $};
		\node[left=-1.3ex of XY] {$\bwdmindepth =\ $};
		\foreach \x [count=\i from 0, count=\j from 1] in \bmArray{
			\node[right, right of=bwdbm\i,xshift=-0.2ex] (bwdbm\j) {\x};
			\ifnum \i > 0 
				\pgfmathparse{ int( {\bmArray}[\i] && ({\bmArray}[int(\i-1)] == 0) }\let\ENTRIES\pgfmathresult
				\ifnum \ENTRIES = 1 
					\pgfmathparse{{\minDepthArray}[int(\theminPosCnt)]}\let\minDepth\pgfmathresult
					\node[text height=2ex, text depth=1ex] at (bwdbm\j |- minDepth0) {\minDepth};
					\addtocounter{minPosCnt}{1}
				\fi	
			\fi	
		}
	\end{scope}
\end{tikzpicture}%
}

\newcommand{\printBlCBWTBmMinDepth}[0]{%
\begin{tikzpicture}[remember picture]
	\begin{scope}[node distance=1.5ex,font=\tt]
		\coordinate (bwdBl0) at (0,0);
		\coordinate (bwdBWT0) at (0,-3ex);
		\node[left=-1.3ex of bwdBl0] {$\bwdbl =\ $};
		\node[left=-1.3ex of bwdBWT0] {$\cL =\ $};
		\foreach \x [count=\i from 0, count=\j from 1] in \bwdBlArray{
			\node[right, right of=bwdBl\i, xshift=-0.2ex] (bwdBl\j) {\x};
			\ifnum \x = 1
				\pgfmathparse{{\bwdBWTArray}[\i]}\let\bwt\pgfmathresult
				\node[text height=2ex, text depth=1ex] at (bwdBl\j |- bwdBWT0) {\bwt};
			\fi	
		}
	\setcounter{minPosCnt}{0}
		\coordinate (bwdbm0) at (0,-8ex);
		\coordinate (minDepth0) at (0,-11ex);
		\coordinate (XY) at (bwdbm0 |- minDepth0);
		\node[left=-1.3ex of bwdbm0] {$\bwdbm =\ $};
		\node[left=-1.3ex of XY] {$\bwdmindepth =\ $};
		\foreach \x [count=\i from 0, count=\j from 1] in \bmArray{
			\node[right, right of=bwdbm\i,xshift=-0.2ex] (bwdbm\j) {\x};
			\ifnum \i > 0 
				\pgfmathparse{ int( ({\bmArray}[\i]==1)  * ({\bmArray}[int(\i-1)] == 0 )) }\let\ENTRIES\pgfmathresult
				\ifnum \ENTRIES = 1 
					\pgfmathparse{{\minDepthArray}[int(\theminPosCnt)]}\let\minDepth\pgfmathresult
					\node[text height=2ex, text depth=1ex] at (bwdbm\j |- minDepth0) {\minDepth};
					\addtocounter{minPosCnt}{1}
				\fi	
			\fi	
		}
	\end{scope}
\end{tikzpicture}%
}

\tikzstyle{st_block_info} = []
\tikzstyle{st_block_fwd_id_label} = [shape=circle,scale=0.6,fill=black, draw=white,%
                                 inner sep=0mm, minimum size=5mm, font=\bf, text=white]

\tikzstyle{st_block_bwd_id_label} = [scale=0.7,above right,inner sep=0mm,text=gray] 
\tikzstyle{st_line_to_block_fwd_id_label} = [densely dashed]
\tikzstyle{st_mark_interval_right} = []
\tikzstyle{st_i} = [xshift=0cm]
\tikzstyle{st_elem_i} = [left]
\tikzstyle{st_sa} = [xshift=2cm]
\tikzstyle{st_elem_sa} = [left]
\tikzstyle{st_lcp} = [xshift=3.2cm]
\tikzstyle{st_elem_lcp} = [left]
\tikzstyle{st_bwt} = [xshift=4.2cm]
\tikzstyle{st_elem_bwt} = [left]
\tikzstyle{st_fwdbf} = [xshift=5.2cm]
\tikzstyle{st_elem_fwdbf} = [left]
\tikzstyle{st_text} = [xshift=5.6cm]
\tikzstyle{st_elem_text} = [right]
\tikzstyle{st_singleton} = [thick]
\tikzstyle{st_reducible} = [thick, fill=black, opacity=0.2]
\tikzstyle{st_irreducible} = [thick, fill=black, opacity=0.6]
\tikzstyle{st_interval} = [thick, fill=orange, fill opacity=0.3]
\tikzstyle{st_fringe} = [draw=white, fill=blue, fill opacity=0.2]
\tikzstyle{st_right_fringe} = 	%
								[st_fringe]
\tikzstyle{st_simple_fringe} = 	[opacity=0]
\tikzstyle{st_left_fringe} = 	%
								[st_fringe]

\tikzstyle{st_full_fringe} = 	%
								[st_fringe]

\tikzstyle{st_interval_edge} = [thick, ->, >=stealth]
\tikzstyle{st_reducible_edge} = [very thick, ->, >=stealth']
\tikzstyle{st_fwd_index_header} = [yshift=0.6cm]
\tikzstyle{st_bwd_index_header} = [yshift=0.6cm]
\tikzstyle{st_y} = [node distance=2.6ex]
\tikzstyle{st_block_header} = [text=red, scale=0.7]

\newcommand{\bitvec}[1]{\ensuremath{\mathsf{#1}}}
\newcommand{\bwdbf}[0]{\bitvec{bf}} %
\newcommand{\bwdbl}[0]{\bitvec{bl}} %
\newcommand{\bwdbm}[0]{\bitvec{bm}} %
\newcommand{\bwdmindepth}[0]{\ensuremath{\mathit{min\mathunderscore depth}}} %
\newcommand{\deltax}[0]{\ensuremath{\Delta_{x}}} %
\newcommand{\deltad}[0]{\ensuremath{\Delta_{d}}} %
\newcommand{\bwdID}[0]{\ensuremath{\mathit{bwd\mathunderscore id}}} %
\newcommand{\bwdid}[1]{\ensuremath{\bwdID(#1)}} %
\newcommand{\runnr}[0]{\ensuremath{run\mathunderscore nr}} %
\newcommand{\runpos}[0]{\ensuremath{run\mathunderscore pos}} %
\newcommand{\access}[0]{\ensuremath{\mathit{access}}}       
\newcommand{\select}[0]{\ensuremath{\mathit{select}}}       
\newcommand{\rank}[0]{\ensuremath{\mathit{rank}}}    

\newcommand{\lb}[0]{\ensuremath{\mathit{lb}}}         %
\newcommand{\rb}[0]{\ensuremath{\mathit{rb}}}         %
\newcommand{\Keyw}[1]{{\bf #1}}

\newcommand{\BWT}[0]{\ensuremath{\mathsf{L}}}
\newcommand{\cBWT}[0]{\ensuremath{\BWT^{\TEXT^{r}}}}
\newcommand{\cL}[0]{\ensuremath{\mathsf{CL}}}
\newcommand{\cC}[0]{\ensuremath{\mathsf{CC}}}

\newcommand{\LCP}[0]{\ensuremath{\mathsf{LCP}}}
\newcommand{\SUF}[0]{\ensuremath{\mathsf{SA}}}

\newcommand{\CARRAY}[0]{\ensuremath{\mathsf{C}}}

\newcommand{\TEXT}[0]{\ensuremath{\mathsf{T}}}
\newcommand{\PATT}[0]{\ensuremath{\mathsf{P}}}
\newcommand{\Order}[1]{\ensuremath{{O}(#1)}}
\newcommand{\order}[1]{\ensuremath{o(#1)}}

\newcommand{\bsstate}[4]{(#1,#2)[#3..#4]} %

\newcommand{\method}[1]{{\sc{#1}}}
\newcommand{\augmented}{\method{Augmented-SA}}
\newcommand{\spat}{\method{SPAT}}
\newcommand{\lofsa}{\method{LOF-SA}}

\newcommand{\rosa}{\method{RoSA}}
\newcommand{\fmindex}{\method{FM-Index}}
\newcommand{\sbtree}{\method{SB-Tree}}
\newcommand{\femto}{\method{FEMTO}}

\newcommand{\bigfile}{\method{Web-64000}}
\newcommand{\smlfile}{\method{Web-4000}}
\newcommand{\tnyfile}{\method{Web-256}}
\newcommand{\dnafile}{\method{DNA-3000}}
\newcommand{\dblpfile}{\method{DBLP-1000}}
\newcommand{\qmode}[1]{{\emph{#1}}}
\newcommand{\qexistence}{\qmode{existence}}
\newcommand{\qcount}{\qmode{count}}
\newcommand{\qlocate}{\qmode{locate}}
\newcommand{\qcontext}{\qmode{context}}
\newcommand{\qstring}[1]{\mbox{``{\tt{#1}}''}}

\newcommand{\size}{\var{size}}
\newcommand{\pparent}{\var{parent}}
\newcommand{\D}{\hphantom{0}}
\newcommand{\C}{\hphantom{0,}}

\newcommand{\mb}[1]{{$#1$~MB}}
\newcommand{\gb}[1]{{$#1$~GB}}
\newcommand{\var}[1]{{\mbox{$\mathit{#1}$}}}
\newcommand{\myparagraph}[1]{\subsection{#1}}

\newcommand{\sdslds}[1]{{\tt{#1}}}
\newcommand{\sdslbitvector}{\sdslds{bit\_vector}}
\newcommand{\sdslsdvector}{\sdslds{sd\_vector<>}}
\newcommand{\sdslrrrvectorvar}{\sdslds{rrr\_vector<63>}}

\begin{document}
\title{Large-Scale Pattern Search Using\\
Reduced-Space On-Disk Suffix Arrays}

\author{Simon~Gog,
        Alistair~Moffat,
        J.~Shane~Culpepper,
        Andrew Turpin,
        and~Anthony~Wirth
\thanks{S. Gog, A. Moffat, A. Turpin, and A. Wirth are with the Department
of Computing and Information Systems, The University of Melbourne, Australia 3010,
e-mail: (see http://www.csse.unimelb.edu.au/\char126alistair/).}%
\thanks{J. S. Culpepper is with the School of Computer Science and Information Technology, RMIT University, Australia 3001.}%
\thanks{Manuscript received March 2013.}}

\markboth{Gog \MakeLowercase{\textit{et al.}}: Reduced-Space On-Disk Suffix Arrays}%
{Gog \MakeLowercase{\textit{et al.}}: Reduced-Space On-Disk Suffix Arrays}

\maketitle

\begin{abstract}
The suffix array is an efficient data structure for in-memory pattern
search.
Suffix arrays can also be used for external-memory pattern search,
via two-level structures that use an internal index to identify the
correct block of suffix pointers.
In this paper we describe a new two-level suffix array-based index
structure that requires significantly less disk space than previous
approaches.
Key to the saving is the use of disk blocks that are based on
prefixes rather than the more usual uniform-sampling approach,
allowing reductions between blocks and subparts of other blocks.
We also describe a new in-memory structure based on a condensed BWT
string, and show that it allows common patterns to be resolved
without access to the text.
Experiments using {\gb{64}} of English web text and a laptop computer
with just {\gb{4}} of main memory demonstrate the speed and
versatility of the new approach.
For this data the index is around one-third the size of previous
two-level mechanisms; and the memory footprint of as little as $1$\%
of the text size means that queries can be processed more quickly
than is possible with a compact {\fmindex}.
\end{abstract}


\IEEEpeerreviewmaketitle

\section{Introduction}
\IEEEPARstart{S}{tring}
search is a well known problem: given a text
$\TEXT[0\ldots n-1]$ over some alphabet $\Sigma$ of size
$\sigma=|\Sigma|$, and a pattern $\PATT[0\ldots m-1]$, locate the
occurrences of $\PATT$ in $\TEXT$.
Several different query modes are possible: asking whether or not
$\PATT$ occurs ({\qexistence} queries); asking how many times
$\PATT$ occurs ({\qcount} queries); asking for the byte locations in
$\TEXT$ at which $\PATT$ occurs ({\qlocate} queries); and asking for
a set of extracted contexts of $\TEXT$ that includes each occurrence
of $\PATT$ ({\qcontext} queries).

When $\TEXT$ and $\PATT$ are provided on a one-off basis, sequential
pattern search methods take $\Order{n+m}$
time.
When $\TEXT$ is fixed, and many patterns are to be processed, it is
likely to be more efficient to pre-process $\TEXT$ and construct an
{\emph{index}}.
The {\emph{suffix array}} {\cite{mm93siamjc}} is one such index,
allowing {\qlocate} queries to be answered in $\Order{m+\log n+k}$
time when there are $k$ occurrences of $\PATT$ in $\TEXT$, using
$\Order{n \log n}$ bits of space in addition to $\TEXT$.
Further alternatives are discussed in Section~\ref{sec-previous}.

Suffix arrays only provide efficient querying if $\TEXT$ plus the
index require less main memory than is available on the host
computer, because random accesses are required to the index and the
text.
For large texts, two-tier structures are needed, with an in-memory
component consulted first in order to identify the data that must be
retrieved from an on-disk index.

\myparagraph{Our Contributions}

We show that if the usual fixed-interval sampling approach to
creating the in-memory index for a two-level suffix array is replaced
by a sampling method that respects common prefixes, the space
required by the suffix array blocks on disk can be reduced by as
much as $50$\%.
This gain is achieved by identifying
{\emph{reducible}} blocks that can be replaced by references to
subintervals within other on-disk blocks.

\begin{figure*}[t]
\rule{0.0mm}{0mm}
\begin{center}
\begin{tabular}{@{}p{9.5cm}|@{\quad\quad}p{5cm}@{}}
\vspace{0pt}
\tikzstyle{st_block_header} = [transparent] 
\tikzstyle{st_interval_edge} = [transparent]
\tikzstyle{st_left_fringe} = [transparent]
\tikzstyle{st_right_fringe} = [transparent]                                
\tikzstyle{st_full_fringe} = [transparent]
\tikzstyle{st_elem_fwdbf} = [transparent]                                  
\tikzstyle{st_text} = [xshift=5.0cm]
\mbox{%
\begin{tikzpicture}[font=\tt, inner sep=0.1mm, remember picture]%

\begin{scope}[st_y]

\coordinate[] (y0) at (0,0);%
\coordinate[below of=y0] (y1);%
\coordinate[below of=y1] (y2);%
\coordinate[below of=y2] (y3);%
\coordinate[below of=y3] (y4);%
\coordinate[below of=y4] (y5);%
\coordinate[below of=y5] (y6);%
\coordinate[below of=y6] (y7);%
\coordinate[below of=y7] (y8);%
\coordinate[below of=y8] (y9);%
\coordinate[below of=y9] (y10);%
\coordinate[below of=y10] (y11);%
\coordinate[below of=y11] (y12);%
\coordinate[below of=y12] (y13);%
\coordinate[below of=y13] (y14);%
\coordinate[below of=y14] (y15);%
\coordinate[below of=y15] (y16);%
\coordinate[below of=y16] (y17);%
\end{scope}%

\node[st_i] (i) at (0,0cm) {};%
\node[st_elem_i] (i0) at (i|- y0) {0};%
\node[st_elem_i] (i1) at (i|- y1) {1};%
\node[st_elem_i] (i2) at (i|- y2) {2};%
\node[st_elem_i] (i3) at (i|- y3) {3};%
\node[st_elem_i] (i4) at (i|- y4) {4};%
\node[st_elem_i] (i5) at (i|- y5) {5};%
\node[st_elem_i] (i6) at (i|- y6) {6};%
\node[st_elem_i] (i7) at (i|- y7) {7};%
\node[st_elem_i] (i8) at (i|- y8) {8};%
\node[st_elem_i] (i9) at (i|- y9) {9};%
\node[st_elem_i] (i10) at (i|- y10) {10};%
\node[st_elem_i] (i11) at (i|- y11) {11};%
\node[st_elem_i] (i12) at (i|- y12) {12};%
\node[st_elem_i] (i13) at (i|- y13) {13};%
\node[st_elem_i] (i14) at (i|- y14) {14};%
\node[st_elem_i] (i15) at (i|- y15) {15};%
\node[st_elem_i] (i16) at (i|- y16) {16};%
\node[st_sa] (sa) at (0,0cm) {};%
\node[st_elem_sa] (sa0) at (sa|- y0) {16};%
\node[st_elem_sa] (sa1) at (sa|- y1) {3};%
\node[st_elem_sa] (sa2) at (sa|- y2) {9};%
\node[st_elem_sa] (sa3) at (sa|- y3) {2};%
\node[st_elem_sa] (sa4) at (sa|- y4) {12};%
\node[st_elem_sa] (sa5) at (sa|- y5) {5};%
\node[st_elem_sa] (sa6) at (sa|- y6) {1};%
\node[st_elem_sa] (sa7) at (sa|- y7) {11};%
\node[st_elem_sa] (sa8) at (sa|- y8) {13};%
\node[st_elem_sa] (sa9) at (sa|- y9) {6};%
\node[st_elem_sa] (sa10) at (sa|- y10) {14};%
\node[st_elem_sa] (sa11) at (sa|- y11) {7};%
\node[st_elem_sa] (sa12) at (sa|- y12) {15};%
\node[st_elem_sa] (sa13) at (sa|- y13) {8};%
\node[st_elem_sa] (sa14) at (sa|- y14) {4};%
\node[st_elem_sa] (sa15) at (sa|- y15) {0};%
\node[st_elem_sa] (sa16) at (sa|- y16) {10};%
\node[st_lcp] (lcp) at (0,0cm) {};%
\node[st_elem_lcp] (lcp0) at (lcp|- y0) {0};%
\node[st_elem_lcp] (lcp1) at (lcp|- y1) {0};%
\node[st_elem_lcp] (lcp2) at (lcp|- y2) {2};%
\node[st_elem_lcp] (lcp3) at (lcp|- y3) {0};%
\node[st_elem_lcp] (lcp4) at (lcp|- y4) {1};%
\node[st_elem_lcp] (lcp5) at (lcp|- y5) {4};%
\node[st_elem_lcp] (lcp6) at (lcp|- y6) {0};%
\node[st_elem_lcp] (lcp7) at (lcp|- y7) {2};%
\node[st_elem_lcp] (lcp8) at (lcp|- y8) {0};%
\node[st_elem_lcp] (lcp9) at (lcp|- y9) {3};%
\node[st_elem_lcp] (lcp10) at (lcp|- y10) {1};%
\node[st_elem_lcp] (lcp11) at (lcp|- y11) {2};%
\node[st_elem_lcp] (lcp12) at (lcp|- y12) {0};%
\node[st_elem_lcp] (lcp13) at (lcp|- y13) {1};%
\node[st_elem_lcp] (lcp14) at (lcp|- y14) {1};%
\node[st_elem_lcp] (lcp15) at (lcp|- y15) {1};%
\node[st_elem_lcp] (lcp16) at (lcp|- y16) {3};%
\node[st_bwt] (bwt) at (0,0cm) {};%
\node[st_elem_bwt] (bwt0) at (bwt|- y0) {s};%
\node[st_elem_bwt] (bwt1) at (bwt|- y1) {e};%
\node[st_elem_bwt] (bwt2) at (bwt|- y2) {s};%
\node[st_elem_bwt] (bwt3) at (bwt|- y3) {h};%
\node[st_elem_bwt] (bwt4) at (bwt|- y4) {h};%
\node[st_elem_bwt] (bwt5) at (bwt|- y5) {s};%
\node[st_elem_bwt] (bwt6) at (bwt|- y6) {s};%
\node[st_elem_bwt] (bwt7) at (bwt|- y7) {s};%
\node[st_elem_bwt] (bwt8) at (bwt|- y8) {e};%
\node[st_elem_bwt] (bwt9) at (bwt|- y9) {e};%
\node[st_elem_bwt] (bwt10) at (bwt|- y10) {l};%
\node[st_elem_bwt] (bwt11) at (bwt|- y11) {l};%
\node[st_elem_bwt] (bwt12) at (bwt|- y12) {l};%
\node[st_elem_bwt] (bwt13) at (bwt|- y13) {l};%
\node[st_elem_bwt] (bwt14) at (bwt|- y14) {\#};%
\node[st_elem_bwt] (bwt15) at (bwt|- y15) {\$};%
\node[st_elem_bwt] (bwt16) at (bwt|- y16) {\#};%
\node[st_fwdbf] (fwdbf) at (0,0cm) {};%
\node[st_elem_fwdbf] (fwdbf0) at (fwdbf|- y0) {1};%
\node[st_elem_fwdbf] (fwdbf1) at (fwdbf|- y1) {1};%
\node[st_elem_fwdbf] (fwdbf2) at (fwdbf|- y2) {0};%
\node[st_elem_fwdbf] (fwdbf3) at (fwdbf|- y3) {1};%
\node[st_elem_fwdbf] (fwdbf4) at (fwdbf|- y4) {0};%
\node[st_elem_fwdbf] (fwdbf5) at (fwdbf|- y5) {0};%
\node[st_elem_fwdbf] (fwdbf6) at (fwdbf|- y6) {1};%
\node[st_elem_fwdbf] (fwdbf7) at (fwdbf|- y7) {0};%
\node[st_elem_fwdbf] (fwdbf8) at (fwdbf|- y8) {1};%
\node[st_elem_fwdbf] (fwdbf9) at (fwdbf|- y9) {0};%
\node[st_elem_fwdbf] (fwdbf10) at (fwdbf|- y10) {1};%
\node[st_elem_fwdbf] (fwdbf11) at (fwdbf|- y11) {0};%
\node[st_elem_fwdbf] (fwdbf12) at (fwdbf|- y12) {1};%
\node[st_elem_fwdbf] (fwdbf13) at (fwdbf|- y13) {1};%
\node[st_elem_fwdbf] (fwdbf14) at (fwdbf|- y14) {1};%
\node[st_elem_fwdbf] (fwdbf15) at (fwdbf|- y15) {1};%
\node[st_elem_fwdbf] (fwdbf16) at (fwdbf|- y16) {0};%
\node[st_elem_fwdbf] (fwdbf17) at (fwdbf|- y17) {1};%
\node[st_text] (text) at (0,0cm) {};%
\node[st_elem_text] (text0) at (text|- y0) {\$};%
\node[st_elem_text] (text1) at (text|- y1) {\#sells\#shells\$};%
\node[st_elem_text] (text2) at (text|- y2) {\#shells\$};%
\node[st_elem_text] (text3) at (text|- y3) {e\#sells\#shells\$};%
\node[st_elem_text] (text4) at (text|- y4) {ells\$};%
\node[st_elem_text] (text5) at (text|- y5) {ells\#shells\$};%
\node[st_elem_text] (text6) at (text|- y6) {he\#sells\#shells\$};%
\node[st_elem_text] (text7) at (text|- y7) {hells\$};%
\node[st_elem_text] (text8) at (text|- y8) {lls\$};%
\node[st_elem_text] (text9) at (text|- y9) {lls\#shells\$};%
\node[st_elem_text] (text10) at (text|- y10) {ls\$};%
\node[st_elem_text] (text11) at (text|- y11) {ls\#shells\$};%
\node[st_elem_text] (text12) at (text|- y12) {s\$};%
\node[st_elem_text] (text13) at (text|- y13) {s\#shells\$};%
\node[st_elem_text] (text14) at (text|- y14) {sells\#shells\$};%
\node[st_elem_text] (text15) at (text|- y15) {she\#sells\#shells\$};%
\node[st_elem_text] (text16) at (text|- y16) {shells\$};%
\end{tikzpicture}%

\markintervalleft{0}{0}{1}{text}{black,st_interval}\markintervalright{0}{0}{2}{i}{st_singleton}
\intervalanchor{0}{0}{i}{block0}
\markintervalleft{1}{2}{1}{text}{black,st_interval}\markintervalright{1}{2}{2}{i}{st_irreducible}
\intervalanchor{1}{2}{i}{block1}
\markintervalleft{3}{5}{1}{text}{black,st_interval}\markintervalright{3}{5}{2}{i}{st_irreducible}
\intervalanchor{3}{5}{i}{block2}
\markintervalleft{6}{7}{1}{text}{black,st_interval}\markintervalright{6}{7}{2}{i}{st_reducible}
\intervalanchor{6}{7}{i}{block3}
\markintervalleft{8}{9}{2}{text}{black,st_interval}\markintervalright{8}{9}{2}{i}{st_reducible}
\intervalanchor{8}{9}{i}{block4}
\markintervalleft{10}{11}{2}{text}{black,st_interval}\markintervalright{10}{11}{2}{i}{st_reducible}
\intervalanchor{10}{11}{i}{block5}
\markintervalleft{12}{12}{2}{text}{black,st_interval}\markintervalright{12}{12}{2}{i}{st_singleton}
\intervalanchor{12}{12}{i}{block6}
\markintervalleft{13}{13}{2}{text}{black,st_interval}\markintervalright{13}{13}{2}{i}{st_singleton}
\intervalanchor{13}{13}{i}{block7}
\markintervalleft{14}{14}{2}{text}{black,st_interval}\markintervalright{14}{14}{2}{i}{st_singleton}
\intervalanchor{14}{14}{i}{block8}
\markintervalleft{15}{16}{2}{text}{black,st_interval}\markintervalright{15}{16}{2}{i}{st_irreducible}
\intervalanchor{15}{16}{i}{block9}
\intervaledge{0}{0}{draw=gray}
\intervaledge{3}{9}{draw=gray}
\intervaledge{4}{2}{draw=gray}
\intervaledge{5}{4}{draw=gray}
\intervaledge{6}{6}{draw=gray}
\intervaledge{7}{7}{draw=gray}
\intervaledge{8}{8}{draw=gray}
\blockinfo{0}{0}{0}{0}{0}
\blockinfo{1}{0}{0}{1}{2}
\blockinfo{2}{0}{0}{2}{4}
\intervaledgereducible{3}{9}{0}
\blockinfo{3}{0}{1}{9}{6}
\intervaledgereducible{4}{2}{1}
\blockinfo{4}{1}{1}{2}{8}
\intervaledgereducible{5}{2}{1}
\blockinfo{5}{1}{2}{2}{9}
\blockinfo{6}{0}{0}{6}{1}
\blockinfo{7}{0}{0}{7}{3}
\blockinfo{8}{0}{0}{8}{5}
\blockinfo{9}{0}{0}{9}{7}
\externalblockheader{1}{{{2,0,0}
}}%
\externalblockheader{2}{{{4,0,0}
,{8,1,1}
,{9,1,2}
}}%
\externalblockheader{9}{{{6,0,1}
,{7,0,0}
}}%
}%
&
\vspace{0pt}
\drawheaderfwd%
\legendforblocksV
\end{tabular}
\end{center}
\mycaption{External common-prefix suffix blocks formed for
$\TEXT={}${\qstring{she\#sells\#shells\$}} with blocksize $b=3$.
\label{fig-rosa-forward}
}
\end{figure*}

We also describe a new in-memory structure for indexing
variable-length common-prefix blocks that is comparable in size to
the {\emph{bit-blind tree}}.
In terms of operational functionality, the new structure has the
benefit of being comprehensive, meaning that {\qexistence} and
{\qcount} searches for frequently-occurring patterns can be resolved
without disk accesses.
The new approach employs backward searching and the Burrows-Wheeler
Transform.

The methodology developed in order to carry out the experimentation
allows independent and stratified exploration of patterns according
to their length and their frequency, and is a third key contribution
of this paper.
Experiments using {\gb{64}} of English web text and a laptop computer
with just {\gb{4}} of main memory demonstrate the speed and
versatility of the new {\rosa} structure.
For this data the {\rosa}'s disk index is around one third of the
size of the previous {\lofsa} two-level suffix-array mechanism
{\cite{spmt08sigmod,mps09dasfaa}}, and the small footprint of the
in-memory part of the index -- as little as $1$\% of the size of the
input text -- means that queries are processed more quickly than is
possible using an {\fmindex}.
That is, while the {\fmindex} {\cite{fm00focs,fm05jacm}} is a much more compact structure, all
of it must be memory-resident during query processing, hindering its
ability to search very large texts.

\myparagraph{Definitions}

Text $\TEXT[0\ldots n-1]$ is assumed to consist of $n$ symbols each a
member of an alphabet $\Sigma=\{a_0, a_1, a_2, \ldots,
a_{\sigma-1}\}$ of size $\sigma=|\Sigma|$, augmented by a sentinel in
$\TEXT[n]$ that is smaller than every element in $\Sigma$.
The $i$\,th suffix of $\TEXT$ is the sequence $\TEXT[i\ldots n]$,
including the sentinel, and is denoted by $\TEXT_i$.
The longest common prefix $\var{LCP}(\TEXT_i, \TEXT_j)$ of two
suffixes of $\TEXT$ is the maximal value $k$ such that
$\TEXT[i+\ell]=\TEXT[j+\ell]$ for all $0\leq\ell < k$.
If $\TEXT_i$ and $\TEXT_j$ are suffixes of $\TEXT$, then
$\TEXT_i<\TEXT_j$ if and only if $\TEXT[i+k]<\TEXT[j+k]$, where
$k=\var{LCP}(\TEXT_i, \TEXT_j)$.
A pattern $\PATT[0\ldots m-1]$ matches $\TEXT$ at $i$ if
$\PATT[0\ldots m-1]$ is identical to $\TEXT[i\ldots i+m-1]$, that is,
if $\PATT$ is a prefix of the $i$\,th suffix of $\TEXT$.

Array $\SUF[0\ldots n]$ is a suffix array for text $\TEXT$ if
$\TEXT_{\SUF[i]} < \TEXT_{\SUF[j]}$ whenever $i<j$.
In the context of a suffix array it is then useful to define
$\LCP[i]=\var{LCP}(\TEXT_{\SUF[i-1]},
\TEXT_{\SUF[i]})$, with $\LCP[0]=-1$.
The Burrows-Wheeler transform (BWT), denoted $\BWT$, is also required
in our development: $\BWT[i]$ contains the preceding character of the
$i$\,th sorted suffix, $\BWT[i]=\TEXT[(\SUF[i]-1)\bmod n]$.
Figure~\ref{fig-rosa-forward} shows an example string of $n=16$
characters that is used throughout the discussion, plus its sorted
suffixes.
The column headed $\SUF[i]$ is the value stored in the $i$\,th entry
in the suffix array for the string; and the column headed $\BWT[i]$
is the corresponding BWT symbol, being the character immediately
prior to the $i$\,th sorted suffix.
The other parts of Figure~\ref{fig-rosa-forward} are described
shortly.

We also employ {\emph{rank}} and {\emph{select}} operations: for
sequence $X$ operation $\rank(X,i,c)$ returns the number of
occurrences of symbol or sequence $c$ in $X[0..i-1]$; and
$\select(X,i,c)$ returns the position of the $i$\,th occurrence of
$c$, counting from zero.
For example, if $X[0..15]=\qstring{she\#sells\#shells}$, then
$\rank(X, 8, {\qstring{s}})$ is $2$, and $\select(X, 2,
{\qstring{e}})$ is $12$.
Although sophisticated mechanisms exist for implementing {\rank} and
{\select} that have good asymptotic properties, one of the most
useful practical approaches simply adds regular cumulative sums to a
standard bitvector representation, expanding it by $25$\% or by
$6.25$\%, depending on the sampling interval~\cite{gp13spe,vigna08wea}.

\section{Background: Suffix Tries, Trees, and Arrays}
\label{sec-previous}

A number of index structures can be used for string search over a
static text $\TEXT$ if it is assumed that $\TEXT$ and its index can
both be held in fast random-access memory.

\myparagraph{Suffix Trie}
\label{subsec-suffixtrie}

A trie is a tree in which each node is implicitly labeled with the
concatenation of the edge labels on the path from the root, and each
of the as many as $\sigma$ edges out of each node is explicitly
labeled with a single symbol from the alphabet $\Sigma$.
A suffix trie for a text $\TEXT$ contains a leaf for each of the
$n+1$ suffixes $\TEXT_i$, each of which stores the corresponding
index $i$.
The storage required by a suffix trie is proportional to the total
number of edges in the trie,
and might be as large as $\Theta(n^2)$.
The minimum space required by a suffix trie is at least $n \log n$
bits\footnote{We assume throughout that
logarithms are binary, and that $\log x$ should be taken to mean
$\lceil \log_2x \rceil$ when appropriate.},
since every location in $\TEXT$ is indexed in the tree, and
involves an address in the range $0\ldots n$.
If the set of child pointers at each node is stored as a table
indexed by edge label, {\qexistence} and {\qcount} queries for a
pattern $\PATT$ of length $m$ can be processed in $\Order{m}$ time,
and {\qlocate} queries in $\Order{m+k}$ time, where $k$ is the number
of matching positions.

\myparagraph{Suffix Tree}

A suffix tree for text $\TEXT$ is a modified suffix trie in which the
parent-child edges represent sequences of symbols from $\Sigma$
rather than single symbols; and in which internal nodes that only
have a single child are eliminated.
The edge labels are stored as references to $\TEXT$ rather than as
explicit sequences of symbols, and the per-edge space
requirement increases from $\Order{\log\sigma}$ bits to $\Order{\log
n}$ bits.
But the number of edges is bounded, and a suffix tree for $\TEXT$ has
$n$ leaves and at most $n$ internal nodes, and occupies at most
$\Order{n\log n}$ bits in total, with typical implementations
requiring $3n$ or more $\log n$-bit pointers.
Searching follows the same process as in a suffix trie, but involves
an access to $\TEXT$ as each edge is traversed, in order to match
symbols in the pattern.

\begin{figure*}[t]
\begin{center}
\includegraphics[scale=0.9]{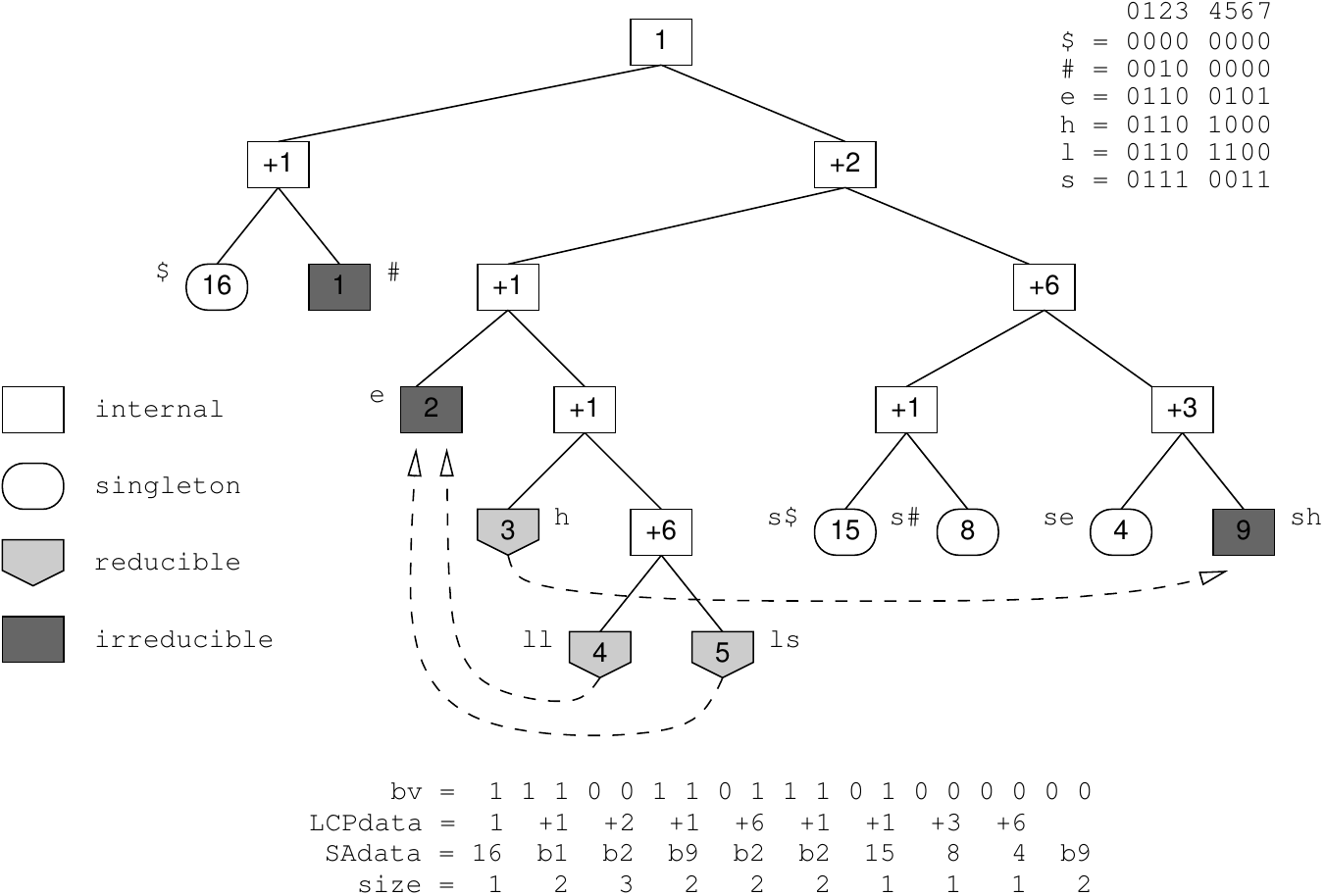}
\end{center}
\mycaption{Bit-blind tree for the ASCII strings
{\qstring{\$}},
{\qstring{\#}},
{\qstring{e}},
{\qstring{h}},
{\qstring{ll}},
{\qstring{ls}},
{\qstring{s\$}},
{\qstring{s\#}},
{\qstring{se}},
and
{\qstring{sh}}, being the identifying block prefixes of the ten
suffix array blocks identified in Figure~\ref{fig-rosa-forward} when
the example string is processed with $b=3$.
The three different types of leaf nodes, and the meaning of the
dotted lines, are discussed in Section~\ref{sec-reducible}.
The ASCII codes for the characters in question are shown at the
top-right.
\label{fig-bitblind}}
\end{figure*}

\myparagraph{Blind Tree}
\label{subsec-bitblind}

The suffix tree's accesses to the text $\TEXT$ are not localized,
and are relatively costly.
In a {\emph{blind tree}}
{\cite{fg99jacm,kr03chapter,mf04algorithmica}} the outgoing edges at
each node are represented by just the first symbol of the
corresponding sequence, rather than by pointers to $\TEXT$.
The remaining (if any) symbols that label that edge in the
corresponding suffix tree are not stored.
Instead, internal nodes store the LCP of the set of strings
represented at that node, and during querying, when a node is
reached, the search steps forward to the indicated symbol, bypassing
any omitted labels.

Search in a blind tree follows a similar path as in an equivalent
suffix tree.
At any given node, at most one edge can match the next unexamined
symbol in the pattern $\PATT$, and if such an edge exists, the search
proceeds to the indicated child.
The risk in following edges that are labeled by just a single symbol
is that the other symbols that are bypassed may not match between
$\PATT$ and $\TEXT$.
To address that risk, once either the pattern has been exhausted, or
a leaf has been reached, the full pattern is rechecked against the
location in $\TEXT$ indicated by any leaf in the subtree rooted at
that node, to examine the bypassed symbols.
By proceeding with the search based on only partial matches, not only
is there a saving in space, but also the majority of the accesses to
$\TEXT$ are eliminated.
Instead, a sequential examination of symbols at a single candidate
location of $\TEXT$ is undertaken, to either verify that a match has
been correctly identified, or to confirm that there cannot be any
occurrences of $\PATT$ in $\TEXT$.

\myparagraph{Bit-Blind Tree}

A concise form of blind tree has been developed~\cite{fg99jacm}
which, for clarity, we refer to here as a {\emph{bit-blind tree}}.
Rather than character LCP values and character edge labels,
bit-based LCP values and binary edge labels are employed.
Moreover, because internal nodes have exactly two children, the
edge labels do not need to be stored.
The tree becomes deeper by a factor of as much as $\log\sigma$;
on the other hand, it
takes less space.
In total, the cost of a bit-blind tree storing the $n$ suffixes
of a text $\TEXT$ is $n-1$ internal nodes,
each containing a bit-LCP value and two pointers (or equivalent); and
$n$ leaves, each containing a $\log n$-bit suffix pointer.

Figure~\ref{fig-bitblind} shows the bit-blind tree for the set of
blocks identified in the right-hand side of
Figure~\ref{fig-rosa-forward}.
The reason that these particular strings are of interest, and only a
partial tree is stored, is discussed shortly.
The ten strings are each represented by one of the leaves of the
tree; the categorization of those leaves into three types is also
described below.

The bitvector $\bitvec{bv}$ at the bottom of Figure~\ref{fig-bitblind}
describes the structure of the bit-blind tree, and eliminates the
need for explicit pointers at the internal nodes.
To create $\bitvec{bv}$ the nodes of the tree are labeled in row-level
order, and a {\qstring{1}} bit is stored for nodes with (a pair of)
children, and a {\qstring{0}} bit is stored if not.
The {\qstring{1}} bits exactly correspond to the locations at which
relative LCP values are required; conversely, the {\qstring{0}} bits
exactly correspond to the locations at which block pointers are
required.
The required tree navigation operations on internal nodes (that is,
node identifiers $x$ such that $\bitvec{bv}[x]=1$) are then provided via
{\rank} and {\select} operations, as follows:
\begin{itemize}
\itemsep 0mm
\item
	$\var{lchild}(x) \leftarrow 2 \times {\rank}(\bitvec{bv}, x, 1)+1$
\item
	$\var{rchild}(x) \leftarrow 2 \times {\rank}(\bitvec{bv}, x, 1)+2$
\item
	$\LCP[x] \leftarrow \LCP[\var{parent}(x)] + \var{LCPdata}[{\rank}(\bitvec{bv}, x, 1)]$
\end{itemize}
where $\var{LCPdata}$ is a dense array of bit-LCP differentials, as
shown at the bottom of the diagram, and $\LCP[\var{parent}(x)]$ will
have been computed during the previous iteration of the tree
traversal loop.
Details of the three types of leaf node, and of the meaning of the
{\var{SAdata}} and {\var{size}} fields, are given in
Section~\ref{sec-reducible}.

\myparagraph{Suffix Array}

As has already been noted, occurrences in $\TEXT$ of a pattern
$\PATT$ can be identified using a binary search in suffix array
$\SUF$ using $\Order{\log n+m}$ character
comparisons~\cite{mm93siamjc}.
In addition, if an LCP array is provided, the set of all matching
locations of $\PATT$ in $\TEXT$ can be identified in $\Order{1}$ time
each once the first one has been identified.
The suffix array is more compact than any of the suffix trie or
suffix tree-based alternatives, including the bit-blind tree, and is
typically represented as a single $\log n$-bit value for each suffix
of $\TEXT$.

M{\"a}kinen and Navarro~\cite{mn04cpm} note that runs in the BWT
string $\BWT$ can be used to identify suffix pointer indirections
that allow space to be saved.
Gonz{\'a}lez and Navarro~\cite{gn07cpm} extended this work,
recognizing repeated patterns of suffix pointer differences using the
{\method{Re-Pair}} compression technique.
But note that when $\TEXT$ is small enough that it fits into
available memory, the {\fmindex}, described next, is the most
attractive option.
That is, reducing the size of an in-memory suffix array does not
necessarily lead to performance improvements.
In Section~\ref{sec-reducible} we apply similar techniques to
disk-based suffix arrays, where the space reduction achieved does
make a difference.

\myparagraph{FM-Index}

The last decade has seen considerable development in the area of
{\emph{compressed self indexing}}.
Hon~et al.~\cite{hsv10cpm} survey much of this work; perhaps the best
exemplar of the category is the {\fmindex} of Ferragina and
Manzini~\cite{fm00focs,fm05jacm}.
Based around the Burrows-Wheeler transform, the {\fmindex} has a
highly desirable blend of properties -- it allows pattern search in
$\Order{m\log\sigma}$ time; it requires space proportional to
$nH_k(\TEXT)+\sigma^k$, the information content of the original
text\footnote{That is, the number of bits required to store the text
using an order-$k$ statistical context-based compression model,
including an allowance for storing the model parameters.}; and it
allows reconstruction of the text in entirety from the beginning, and
from (with additional storage cost) sampled re-entry points.

For texts for which the {\fmindex} fits into random access memory,
{\qexistence} and {\qcount} queries are fast; while
the speed of {\qlocate} queries depends on the sampling rate for
decoding, and allows a tradeoff between space and speed.
We include experimental results for the {\fmindex} in
Section~\ref{sec-experiments}, based on a new implementation
developed as part of a recent investigation into compressed
bitvector representations~\cite{gp13spe}.

The {\fmindex} is less efficient when the compressed representation
of $\TEXT$ is too large for main memory -- the non-sequential access
pattern dictated by the BWT sequence makes the {\fmindex} a poor
choice for disk-based search.
A particular disadvantage of the backward search used in the
{\fmindex} is that the range of the search interval is
non-increasing, but the upper and lower bounds on that interval are
not convergent.
This arrangement means that even the best external variants of
compressed searching potentially make $m$ disk
accesses~\cite{gn09jcmcc}, which is impractical for long
patterns.

\section{On-Disk Suffix Arrays}
\label{sec-ondisksuffix}

Two approaches have emerged for storing suffix array structures on
secondary storage: methods that make use of uniform-size blocks, so
that every block except the last contains exactly $b$ pointers; and
methods that make use of variable-sized blocks, in which $b$ is an
upper bound on the blocksize, and characteristics of the data are
used to determine the block boundaries, subject to that bound.

\myparagraph{Uniform Blocks and the String B-Tree}

Baeza-Yates et al.~\cite{bybz96infsys} describe the {\spat}, a
structure in which the suffix array is formed into uniform blocks
each containing $b$ pointers, and the in-memory index is an array of
$n/b$ fixed-length strings, being the first $\ell_s$ symbols of the
last suffix in each block.
The {\augmented} proposal of Colussi and De~Col~\cite{cd96ipl} also
partitions the on-disk suffix array into uniform blocks (each of
$b=\log n$ suffix pointers)
but with the in-memory index
constructed as a suffix tree to the (full) first suffix string of the
block.
Gonz{\'a}lez and Navarro~\cite{gn09jcmcc} provide a summary of these
early techniques.

Ferragina and Grossi~\cite{fg99jacm} describe
a dynamic string search structure they call the String B-tree, or
{\sbtree}.
For static data of the type considered here, the {\sbtree} is
implemented as a uniform partitioning of a suffix array, with an
in-memory suffix tree index implemented as a blind tree or bit-blind
tree, with each leaf containing a block of $b$ suffix pointers to
$\TEXT$.
More than one level of indexing can be used if necessary, with all
blocks having the same structure.
Each node of the {\sbtree} indexes $b$ strings via $2b$ bits
describing the shape of a binary tree of $b$ leaves and $b-1$
internal nodes; plus $b-1$ internal node depths, expressed in bit
offsets from the start of the pattern $\PATT$, each taking at most
$\log(\hat{n}\log\sigma)$ bits, where $\hat{n}\le n$ is the longest
character LCP value across the entire set of strings; plus $b$ suffix
pointers each of $\log n$ bits.
If all non-leaf blocks are held in memory, the in-memory component of
a static {\sbtree} contains $s=\lceil n/b\rceil$ suffix pointers,
$s-1$ LCP values, and $2s$ bits describing the tree shape.

An advantage of the uniform-sized disk blocks used in the {\sbtree}
is that they allow node addresses to be calculated rather than
stored, and no pointers are needed to navigate the index.
The only pointers stored in the {\sbtree} -- in internal nodes as
well as in leaf blocks -- are to the text $\TEXT$ rather than to disk
blocks.
Note also that suffix pointers are required in the internal nodes for
a static {\sbtree} only if blind search-induced pattern ambiguity is
to be resolved on a per-block basis.
If the pattern ambiguity is tolerated until the whole of $\PATT$ has
been handled, then a single holistic check can be undertaken against
$\TEXT$ via a suffix pointer from a leaf node.
Regardless, as a minimum, a static {\sbtree} stores $n$ suffix
pointers in its leaf blocks, occupying $n \log n$ bits.

Taking these various considerations into account, the minimum size
for a static {\sbtree} covering a text of $n$ symbols using a
blocksize of $b$ pointers is
\begin{equation}
	n \left( 2 + \log(\hat{n}\log\sigma) + \log n \right)
\label{eqn-sbtree}
\end{equation}
bits where $\hat{n}< n $ is the length in characters of the largest
LCP value.
That is, the {\sbtree} index might add a space overhead of as much as
$100$\% to the $n \log n$ bits required by a plain suffix array.

\myparagraph{Variable Blocks and the LOF-SA}

Sinha et al.~\cite{spmt08sigmod} describe the {\lofsa}, a two-level
index structure in which the block control parameter $b$ is an upper
bound, and suffix array blocks
correspond to subtrees in the suffix tree.
If $v$ is a node in the suffix tree for text $\TEXT$, and if
$\size(\pparent(v)) >b $ and $\size(v)\le b$, then a suffix array
block is formed corresponding to node $v$.
All elements in the block share the prefix associated with $v$.
The divisions shown in Figure~\ref{fig-rosa-forward} denote the ten
blocks that result when the example string is processed allowing at
most $b=3$ suffix pointers in each suffix block; and
Figure~\ref{fig-bitblind} shows how those ten block prefixes are
stored in a bit-blind tree.

Sinha~et al.~use a trie for the in-memory component of the {\lofsa},
but this has the disadvantage of a quadratic worst-case space
requirement.
A bit-blind trie, and the structure we present in the
Section~\ref{sec-condensed}, both require less space in both the
average case and the worst case.

Pattern search using the {\lofsa} steps through the symbols in
$\PATT$, navigating the in-memory search structure, either until the
pattern is exhausted, in which case all children of the node that was
reached are answers to the query; or until a leaf in the trie is
reached, in which case the answers, if any exist, are confined to a
single block of the on-disk suffix array.
In the latter case that block is fetched and searched.

Regardless of how the internal structure is organized, the variable
sized disk blocks mean that a disk address of $\log n$ bits must be
stored at each in-memory leaf.
In the on-disk blocks, Sinha et al.~also store an LCP value for each
suffix; plus, as was previously sketched by Colussi and
De~Col~\cite{cd96ipl}, a small number $f$ of extension symbols (the
{\emph{fringe}}) to help minimize search ambiguities.
Search within a {\lofsa} suffix block is sequential, capitalizing on
the LCP and fringe values.
Accesses are made to $\TEXT$ only if there are gaps in the fringe
that result in pattern uncertainty.
Inclusion of the fringe for each suffix increases the size of disk
blocks, and each entry in each on-disk suffix block contains an
$\LCP$ value, a pointer into $\TEXT$, and a set of fringe symbols.

Sinha et al.~undertook a range of experiments with {\gb{2}} of DNA
and {\mb{471}} of English text, and patterns of length $6$ to
$1{,}000$.
With a blocksize bound of $b=4{,}096$ and a fringe length of $f=4$
characters, the in-memory component and on-disk component for the
{\mb{471}} English text file required {\mb{21}} and {\gb{5.5}}
respectively, and yielded searching times around half or less of the
{\spat}, and around $8$ times faster than a pure suffix array.
Moffat et al.~\cite{mps09dasfaa} considered
compression of the on-disk components, and showed that the space
required by the on-disk data can be reduced by approximately $40$\%,
from $12n$ bytes down to around $7.1n$ bytes.

The next two sections describe our enhancements to the {\lofsa}.
First, in Section~\ref{sec-reducible}, we show that as many as half
of the suffix pointers can be elided, via a process we call
{\emph{block reduction}}.
Then, in Section~\ref{sec-condensed} we introduce a {\emph{condensed
BWT}} in-memory index structure that provides a unique mix of
attributes and allows fast searching over a set of strings.

\section{Reducible Blocks}
\label{sec-reducible}

This section considers the suffix array reduction process of
M{\"a}kinen and Navarro~\cite{mn04cpm} and shows that it can be
applied to variable-size on-disk suffix blocks.

\myparagraph{Identifying Reductions}

A whole-block reduction is possible exactly when all of the BWT
symbols corresponding to the suffixes contained in a block are the
same.
For example, in Figure~\ref{fig-rosa-forward}, the suffixes
corresponding to the prefix {\qstring{h}}, with pointers $\SUF[6]=1$
and $\SUF[7]=11$, form a block when $b=3$; and both have an
{\qstring{s}} in the column headed $\BWT[i]$.
Hence, a reduction to the suffix {\qstring{sh}} is possible.
Examination of the set of $b=3$ blocks shown in
Figure~\ref{fig-rosa-forward} reveals that the suffixes at offsets
$8$--$9$ for {\qstring{ll}} can be reduced to a (subset of) the block
at suffixes at offset $3$--$5$ for {\qstring{e}}; and that, via two
such steps, the suffixes at offsets $10$--$11$ for {\qstring{ls}} can
be reduced to the same underlying block.
The three arrows at the left of Figure~\ref{fig-rosa-forward} show
the full set of block relationships that exist in the example string,
with the three reducible blocks lightly shaded; the same reductions
are also noted with the dotted arrows in Figure~\ref{fig-bitblind}.

\myparagraph{Singleton Blocks}

The variable block approach also sometime generates blocks with just
one pointer in them; we call these {\emph{singleton blocks}}.
They are unshaded in Figures~\ref{fig-rosa-forward}
and~\ref{fig-bitblind}, and represent another opportunity for space
savings, since the corresponding suffix pointers can be stored
directly in the in-memory index, rather than placed in a suffix block
on disk.
In the example string there are four singleton blocks.
Only non-singleton {\emph{irreducible blocks}} need to be placed onto
disk; as can be seen in the example, there are three such blocks, and
they contain a total of only seven suffix pointers.

\myparagraph{Storing Information About Reductions}

The details of each block reduction are held as a
$(\Delta_x,\Delta_d)$ pair relative to an irreducible block, where
$\Delta_x$ is the offset from the start of the irreducible block at
which the reduced block commences, and $\Delta_d$ is the offset to be
applied to each suffix pointer.
The three reducible blocks in
Figure~\ref{fig-rosa-forward} are annotated with their offset pairs.

To save memory space, each leaf of the in-memory index stores only a
block number, and all other information is stored as part of the disk
blocks.
Each suffix block contains a small header table of $\Delta_x$ and
$\Delta_d$ values, one pair per suffix block (reducible or
irreducible) that is hosted within that set of suffix pointers.
This table maps information accumulated during the in-memory search
(position reached in the pattern, and current suffix interval width)
to $(\Delta_x,\Delta_d)$ pairs that are used to continue the search
within the block.
The in-memory part does not differentiate between reducible and
irreducible blocks at all  -- the latter correspond to
$\Delta_x=0$ and $\Delta_d=0$.

The in-memory structure identifies singletons by virtue of the fact
that the search interval is one.
Singletons are also reducible, by definition, but a search-time disk
access can be saved if they point directly to $\TEXT$ rather than via
a suffix block.
In Figure~\ref{fig-bitblind} non-singleton pointers are marked with a
{\qstring{b}}, but no such differentiation is required in practice,
since singleton-block suffix pointers exactly correspond to
situations where the {\var{size}} field (shown at the bottom of
Figure~\ref{fig-bitblind}) is~$1$.

\myparagraph{Storing the On-Disk Suffix Array}

Each suffix block contains a table of $(\Delta_x,\Delta_d)$ offsets,
plus a set of suffix pointers, plus a set of differential (relative
to the parent) LCP values, plus two bits per leaf to indicate the
tree structure, plus a small fixed overhead on the latter to allow
{\rank} operations.
One key advantage of the {\lofsa} variable-block arrangement
is that each block can store the LCP values (shown as $\var{LCPdata}$
in Figure~\ref{fig-bitblind}) in compressed form, since there is no
requirement that all disk blocks be the same size.
This difference is significant in terms of space utilization.

In our implementation the LCPs are stored as differences relative to
their parent in the suffix tree, and coded using the Elias~$\delta$
code~\cite{wmb99mg} with cumulative-sum samples inserted every $64$
values to allow pseudo-random access to be carried out.
The node sizes are similarly stored cumulatively, so that the size of
any node can be extracted by subtracting the cumulative count of its
leftmost child from the cumulative count of its rightmost child.
Suffix pointers are stored as minimal-width binary values, but are
not otherwise compressed.
We also experimented with an alternative approach, in which LCP
values were stored without being differenced relative to their
parents, and the tree structure was created from the LCP values
rather than via the bitvector {\bitvec{bv}}.
This option turned out to be both larger in size and slower in
operation, and was not pursued beyond preliminary experimentation.

\begin{figure*}[t!]
\begin{center}
\begin{tabular}{@{}p{8.8cm}|@{\quad}p{5cm}@{}}
\vspace{0pt}
\bwdIndexLayout
\mbox{%
\begin{tikzpicture}[font=\tt, inner sep=0.1mm, remember picture]%

\begin{scope}[st_y]

\coordinate[] (y0) at (0,0);%
\coordinate[below of=y0] (y1);%
\coordinate[below of=y1] (y2);%
\coordinate[below of=y2] (y3);%
\coordinate[below of=y3] (y4);%
\coordinate[below of=y4] (y5);%
\coordinate[below of=y5] (y6);%
\coordinate[below of=y6] (y7);%
\coordinate[below of=y7] (y8);%
\coordinate[below of=y8] (y9);%
\coordinate[below of=y9] (y10);%
\coordinate[below of=y10] (y11);%
\coordinate[below of=y11] (y12);%
\coordinate[below of=y12] (y13);%
\coordinate[below of=y13] (y14);%
\coordinate[below of=y14] (y15);%
\coordinate[below of=y15] (y16);%
\coordinate[below of=y16] (y17);%
\end{scope}%

\node[st_i] (i) at (0,0cm) {};%
\node[st_elem_i] (i0) at (i|- y0) {0};%
\node[st_elem_i] (i1) at (i|- y1) {1};%
\node[st_elem_i] (i2) at (i|- y2) {2};%
\node[st_elem_i] (i3) at (i|- y3) {3};%
\node[st_elem_i] (i4) at (i|- y4) {4};%
\node[st_elem_i] (i5) at (i|- y5) {5};%
\node[st_elem_i] (i6) at (i|- y6) {6};%
\node[st_elem_i] (i7) at (i|- y7) {7};%
\node[st_elem_i] (i8) at (i|- y8) {8};%
\node[st_elem_i] (i9) at (i|- y9) {9};%
\node[st_elem_i] (i10) at (i|- y10) {10};%
\node[st_elem_i] (i11) at (i|- y11) {11};%
\node[st_elem_i] (i12) at (i|- y12) {12};%
\node[st_elem_i] (i13) at (i|- y13) {13};%
\node[st_elem_i] (i14) at (i|- y14) {14};%
\node[st_elem_i] (i15) at (i|- y15) {15};%
\node[st_elem_i] (i16) at (i|- y16) {16};%
\node[st_bwt] (bwt) at (0,0cm) {};%
\node[st_elem_bwt] (bwt0) at (bwt|- y0) {s};%
\node[st_elem_bwt] (bwt1) at (bwt|- y1) {s};%
\node[st_elem_bwt] (bwt2) at (bwt|- y2) {s};%
\node[st_elem_bwt] (bwt3) at (bwt|- y3) {\#};%
\node[st_elem_bwt] (bwt4) at (bwt|- y4) {l};%
\node[st_elem_bwt] (bwt5) at (bwt|- y5) {l};%
\node[st_elem_bwt] (bwt6) at (bwt|- y6) {e};%
\node[st_elem_bwt] (bwt7) at (bwt|- y7) {e};%
\node[st_elem_bwt] (bwt8) at (bwt|- y8) {l};%
\node[st_elem_bwt] (bwt9) at (bwt|- y9) {l};%
\node[st_elem_bwt] (bwt10) at (bwt|- y10) {s};%
\node[st_elem_bwt] (bwt11) at (bwt|- y11) {s};%
\node[st_elem_bwt] (bwt12) at (bwt|- y12) {h};%
\node[st_elem_bwt] (bwt13) at (bwt|- y13) {e};%
\node[st_elem_bwt] (bwt14) at (bwt|- y14) {h};%
\node[st_elem_bwt] (bwt15) at (bwt|- y15) {\$};%
\node[st_elem_bwt] (bwt16) at (bwt|- y16) {\#};%
\node[st_bwdbl] (bwdbl) at (0,0cm) {};%
\node[st_elem_bwdbl] (bwdbl0) at (bwdbl|- y0) {1};%
\node[st_elem_bwdbl] (bwdbl1) at (bwdbl|- y1) {0};%
\node[st_elem_bwdbl] (bwdbl2) at (bwdbl|- y2) {0};%
\node[st_elem_bwdbl] (bwdbl3) at (bwdbl|- y3) {1};%
\node[st_elem_bwdbl] (bwdbl4) at (bwdbl|- y4) {1};%
\node[st_elem_bwdbl] (bwdbl5) at (bwdbl|- y5) {0};%
\node[st_elem_bwdbl] (bwdbl6) at (bwdbl|- y6) {1};%
\node[st_elem_bwdbl] (bwdbl7) at (bwdbl|- y7) {0};%
\node[st_elem_bwdbl] (bwdbl8) at (bwdbl|- y8) {1};%
\node[st_elem_bwdbl] (bwdbl9) at (bwdbl|- y9) {0};%
\node[st_elem_bwdbl] (bwdbl10) at (bwdbl|- y10) {1};%
\node[st_elem_bwdbl] (bwdbl11) at (bwdbl|- y11) {0};%
\node[st_elem_bwdbl] (bwdbl12) at (bwdbl|- y12) {1};%
\node[st_elem_bwdbl] (bwdbl13) at (bwdbl|- y13) {1};%
\node[st_elem_bwdbl] (bwdbl14) at (bwdbl|- y14) {0};%
\node[st_elem_bwdbl] (bwdbl15) at (bwdbl|- y15) {1};%
\node[st_elem_bwdbl] (bwdbl16) at (bwdbl|- y16) {1};%
\node[st_bwdbf] (bwdbf) at (0,0cm) {};%
\node[st_elem_bwdbf] (bwdbf0) at (bwdbf|- y0) {1};%
\node[st_elem_bwdbf] (bwdbf1) at (bwdbf|- y1) {1};%
\node[st_elem_bwdbf] (bwdbf2) at (bwdbf|- y2) {1};%
\node[st_elem_bwdbf] (bwdbf3) at (bwdbf|- y3) {1};%
\node[st_elem_bwdbf] (bwdbf4) at (bwdbf|- y4) {0};%
\node[st_elem_bwdbf] (bwdbf5) at (bwdbf|- y5) {1};%
\node[st_elem_bwdbf] (bwdbf6) at (bwdbf|- y6) {1};%
\node[st_elem_bwdbf] (bwdbf7) at (bwdbf|- y7) {0};%
\node[st_elem_bwdbf] (bwdbf8) at (bwdbf|- y8) {1};%
\node[st_elem_bwdbf] (bwdbf9) at (bwdbf|- y9) {0};%
\node[st_elem_bwdbf] (bwdbf10) at (bwdbf|- y10) {1};%
\node[st_elem_bwdbf] (bwdbf11) at (bwdbf|- y11) {0};%
\node[st_elem_bwdbf] (bwdbf12) at (bwdbf|- y12) {1};%
\node[st_elem_bwdbf] (bwdbf13) at (bwdbf|- y13) {0};%
\node[st_elem_bwdbf] (bwdbf14) at (bwdbf|- y14) {0};%
\node[st_elem_bwdbf] (bwdbf15) at (bwdbf|- y15) {1};%
\node[st_elem_bwdbf] (bwdbf16) at (bwdbf|- y16) {0};%
\node[st_elem_bwdbf] (bwdbf17) at (bwdbf|- y17) {1};%
\node[st_text] (text) at (0,0cm) {};%
\node[st_elem_text] (text0) at (text|- y0) {\$};%
\node[st_elem_text] (text1) at (text|- y1) {\#ehs\$};%
\node[st_elem_text] (text2) at (text|- y2) {\#slles\#ehs\$};%
\node[st_elem_text] (text3) at (text|- y3) {ehs\$};%
\node[st_elem_text] (text4) at (text|- y4) {ehs\#slles\#ehs\$};%
\node[st_elem_text] (text5) at (text|- y5) {es\#ehs\$};%
\node[st_elem_text] (text6) at (text|- y6) {hs\$};%
\node[st_elem_text] (text7) at (text|- y7) {hs\#slles\#ehs\$};%
\node[st_elem_text] (text8) at (text|- y8) {lehs\#slles\#ehs\$};%
\node[st_elem_text] (text9) at (text|- y9) {les\#ehs\$};%
\node[st_elem_text] (text10) at (text|- y10) {llehs\#slles\#ehs\$};%
\node[st_elem_text] (text11) at (text|- y11) {lles\#ehs\$};%
\node[st_elem_text] (text12) at (text|- y12) {s\$};%
\node[st_elem_text] (text13) at (text|- y13) {s\#ehs\$};%
\node[st_elem_text] (text14) at (text|- y14) {s\#slles\#ehs\$};%
\node[st_elem_text] (text15) at (text|- y15) {sllehs\#slles\#ehs\$};%
\node[st_elem_text] (text16) at (text|- y16) {slles\#ehs\$};%
\end{tikzpicture}%

\def\fwdBlocksInBwd{%
{0,1,1,0}
,{0,1,2,6}
,{1,3,1,1}
,{2,3,2,7}
,{3,6,1,2}
,{5,6,2,8}
,{6,8,1,3}
,{6,8,2,9}
,{10,12,2,4}
,{15,17,2,5}
}%
\def\bwdBlArray{%
1,0,0,1,1,0,1,0,1,0,1,0,1,1,0,1,1}%
\def\bwdBWTArray{%
"s","s","s","\noexpand\#","l","l","e","e","l","l","s","s","h","e","h","\noexpand\$","\noexpand\#"}%
\def\bmArray{%
0,0,1,0,1,0,1,0,1,0,1,0,0,1,1,0,1,1,0,1,1}%
\def\minDepthArray{%
1,1,2,1,2,1,2,2}%
\drawheaderbwd%
}%
&
\vspace{0pt}
\def\fwdBlocksInBwd{ {0,1,1,0} ,{0,1,2,6} ,{1,3,1,1} ,{2,3,2,7} ,{3,6,1,2} ,{5,6,2,8} ,{6,8,1,3} ,{6,8,2,9} ,{10,12,2,4} ,{15,17,2,5} }%
\def\bwdBlArray{1,0,0,1,1,0,1,0,1,0,1,0,1,1,0,1,1}%
\def\bwdBWTArray{ "s","s","s","\noexpand\#","l","l","e","e","l","l","s","s","h","e","h","\noexpand\$","\noexpand\#"}
\def\bmArray{0,0,1,0,1,0,1,0,1,0,1,0,0,1,1,0,1,1,0,1,1}%
\def\minDepthArray{1,1,2,1,2,1,2,2}%
\markFwdBlocksInBwdIndex
\printBlCBWTBmMinDepth
\end{tabular}
\end{center}
\mycaption{Full BWT text $\cBWT$, condensed BWT text $\cL$, and
indexing bitvectors {\bwdbf} and {\bwdbl} for the reversed text
$\TEXT^r={}${\qstring{sllehs\#slles\#ehs\$}}.
\label{fig-rosa-backward}
}
\end{figure*}

\section{Indexing Using a Condensed BWT}
\label{sec-condensed}

Having devised a mechanism for efficiently determining and storing
block reductions, we now return to the issue of how to provide an
efficient representation of the in-memory index, and introduce a
{\emph{condensed BWT}} index that provides the ability to resolve
{\qexistence} and {\qcount} queries for frequently appearing patterns
(patterns that occur more than $b$ times in~$\TEXT$) without any disk
blocks needing to be retrieved.
The critical observation that makes our approach possible is that
reversing each of the strings stored in the in-memory index allows
backward search within them to match a prefix of the pattern.
Compared to the bit-blind tree, the new approach has the advantage of
being comprehensive, in that the symbols in the pattern are checked
exhaustively.

\myparagraph{Indexing the Blocks}

The {\lofsa} employs a suffix trie (Section~\ref{subsec-suffixtrie})
to store the set of block prefix strings, but requires quadratic
space in the worst-case.
A second option is to use a bit-blind tree
(Section~\ref{subsec-bitblind}).
Figure~\ref{fig-bitblind} shows a tree storing the block prefix
strings for the example text.
Each of the ten leaves corresponds to one of the blocks shown in
Figure~\ref{fig-rosa-forward}; only the irreducible blocks, shown
with dark shading, need to be stored on disk.

When $\bitvec{bv}[x]=0$ and $x$ is the identifier of a leaf, the
quantity $\var{SAdata}[x-{\rank}(\bitvec{bv}, x, 1)]$ indicates where
corresponding suffix pointer(s) are located, with $\var{SAdata}$
another dense array, containing either suffix array pointers, or
suffix block disk addresses (indicated in the example by a
{\qstring{b}} prefix).
The {\var{size}} array also allows {\qcount} queries to be handled
efficiently.

In total, if there are $K$ suffix array blocks, the structure shown
in Figure~\ref{fig-bitblind} requires storage of: $2K$ bits for the
tree structure; $K-1$ bit-LCP differentials, each of which is less
than $n\log\sigma$; $K$ suffix or disk pointers, each of which is
less than $n$; and $K$ block sizes, each of which is less than $b$.
In the worst case, processing of a pattern $\PATT$ of length $m$
requires navigation of the tree from the root to a leaf, and involves
$m\log \sigma$ bit-extraction operations and the same number of rank
operations, and takes $\Order{m\log \sigma}$ time.

\myparagraph{Backward Search in a Forward BWT} 

Ferragina and Manzini~\cite{fm00focs} show that pattern matching can
be realized via the BWT string $\BWT$.
Suppose that a suffix $\omega=\PATT[m\!-\!i..m\!-\!1]$ of length $i$
has been matched, and that the corresponding {\SUF}-interval is
$[\lb_i..\rb_i]$.
We denote this configuration with the notation
$\bsstate{\omega}{i}{\lb_i}{\rb_i}$.
At the beginning of the search, $\bsstate{\epsilon}{0}{0}{n-1}$ is established.
The new {\SUF}-interval $[\lb_{i+1}..\rb_{i+1}]$ for
$\omega'=c\omega$ with $c=P[m\!-i\!-1]$ is contained within the
section of {\SUF} corresponding to strings that commence with $c$.
The offset from the start of that range is computed by counting the
number of length-$i$ substrings which are both lexicographically
smaller than $\omega$ and preceded by $c$.
Hence, $\bsstate{c\omega}{i+1}{\CARRAY[c]
+\rank(\BWT,\lb_i,c)}{\CARRAY[c]+\rank(\BWT,\rb_i+1,c)-1}$ is the
next configuration of the backward search, where $\CARRAY$ is a
$\sigma$-element array that stores in $\CARRAY[c]$ the location in
$\SUF$ of the first suffix commencing with symbol $c$, and can be
computed when $\BWT$ is constructed.

The best approach for {\rank} on general sequences over a non-binary
alphabet is to use a wavelet tree~\cite{ggv03soda} or
variant thereof, which reduces each operation to at most $\log\sigma$
operations over binary sequences.
Here we use a Huffman-shaped tree using compressed
bitvectors~\cite{rrr02soda}, which represents a sequence of symbols
in its $H_0$ self-entropy.
As already noted, on a binary alphabet, {\rank} and {\select} can be
carried out in constant time by adding a fixed overhead ($25$\% or
$6.25$\%) on top of the original bitvector~\cite{gp13spe,vigna08wea}.

\myparagraph{Backward Search in a Condensed Backward BWT} 

A backward search in a reversed text is equivalent to a forward
search in a forward text.
Figure~\ref{fig-rosa-backward} shows the reversed example text in
sorted suffix order, with a number of divisions marked on the
right-hand side.
The column headed $\cBWT[i]$ shows the full BWT of the reversed text;
but for our purposes only a subset of the BWT is required, shown in
the example as $\cL={}${\qstring{s\#lelshe\$\#}}.
To allow positions in the condensed BWT to be mapped to their
positions in $\cBWT$, the bitvector {\bwdbf} is used, with
$\bwdbf[i]=1$ when the predecessor symbol of the $i$\,th suffix is in
$\cL$.
Similarly, bitvector $\bwdbl[i]=1$ if the $i$\,th entry of $\cBWT$
appears in $\cL$.
The run-length compressed {\fmindex} of M\"akinen and
Navarro~\cite{mn05cpm} makes use of auxiliary bitvectors in a similar
manner to what we are about to describe.

Consider the suffix strings on the right-hand side of
Figure~\ref{fig-rosa-forward}.
The block-prefixes (shown by the shading) that need to be reversed
and indexed are
\qstring{\$},
\qstring{\#},
\qstring{e},
\qstring{h},
\qstring{ll},
\qstring{ls},
\qstring{s\$},
\qstring{s\#},
\qstring{se}, and
\qstring{sh}.
When reversed, they become
\qstring{\$},
\qstring{\#},
\qstring{e},
\qstring{h},
\qstring{ll},
\qstring{sl},
\qstring{\$s},
\qstring{\#s},
\qstring{es}, and
\qstring{hs};
if those reversed strings were then formed into a suffix trie, nodes
would be created for all of
\qstring{\$},
\qstring{\$s},
\qstring{\#},
\qstring{\#s},
\qstring{e},
\qstring{es},
\qstring{h},
\qstring{hs},
\qstring{l},
\qstring{ll},
\qstring{s}, and
\qstring{sl}.
To create the bitvector $\bwdbf$ that indicates which of the BWT
characters are needed in the condensed BWT, the interval $[\lb,\rb]$
associated with each of these nominal suffix trie nodes is located in
the reversed BWT, and the bits $\bwdbf[\lb]$ and $\bwdbf[\rb+1]$ are
set to $1$, to mark the beginning and end of each reversed search
interval.
Any locations in $\bwdbf$ with $1$-bits at the end of this stage have
their corresponding first suffix character located in $\cBWT$ and
copied in to $\cL$; and an inverse mapping $\bwdbl$ is computed that
stores the locations extracted.
For example, in Figure~\ref{fig-rosa-backward} the first and fourth
suffixes commencing with {\qstring{s}} are tagged in $\bwdbf$; those
{\qstring{s}} symbols occur in positions $\cBWT[0]$ and $\cBWT[10]$,
and so both $\bwdbl[0]$ and $\bwdbl[10]$ are set to $1$, and two
{\qstring{s}} symbols appear in~$\cL$.
Finally,
set of condensed symbol counts $\cC$ is formed from the
condensed BWT string $\cL$.

\begin{figure}[t]
\begin{tabbing}
\quad \=\quad \=\quad \=\quad \=\quad \=\quad \=\quad \=\quad \=\quad \=\quad \=\quad \=\quad \kill \\
\scriptsize 00 \>\> $\var{get\_interval}(\PATT, m)$\\ 
\scriptsize 01 \>\>\> $d \leftarrow 0$; $\lb \leftarrow 0$;
			$\rb\leftarrow n-1$\\
\scriptsize 02 \>\>\> \Keyw{while} $d < m$ \Keyw{and}
			$\rb - \lb +1 > b$ \Keyw{do} \\
\scriptsize 03 \>\>\>\> $c \leftarrow \PATT[d]$\\
\scriptsize 04 \>\>\>\> $(\lb', \rb') \leftarrow
			( \rank(\bwdbl, \lb, \qstring{1}),
			  \rank(\bwdbl,\rb+1, \qstring{1}) )$\\
\scriptsize 05 \>\>\>\> $(\lb'', \rb'') \leftarrow 
			( \rank(\cL, \lb',c),
				\rank(\cL, \rb',c) )$\\
\scriptsize 06 \>\>\>\> \Keyw{if} $\lb'' = \rb''$ \Keyw{then}\\
\scriptsize 07 \>\>\>\>\> \Keyw{return} $\var{not\_found}$ \\
\scriptsize 08 \>\>\>\> $\lb \leftarrow \select%
			(\bwdbf, \cC[c]+\lb'', \qstring{1})$\\
\scriptsize 09 \>\>\>\> $\rb \leftarrow \select%
			(\bwdbf, \cC[c]+\rb'', \qstring{1})-1$\\
\scriptsize 10 \>\>\>\> $d \leftarrow d + 1$\\
\scriptsize 11 \>\>\> \Keyw{return} $\bsstate{\PATT[0..d-1]}{d}{\lb}{\rb}$
\end{tabbing}
\mycaption{Backward search using a condensed BWT text {\cL} and a
condensed count array $\cC$.
\label{fig-alg-backward}
}
\end{figure}

Figure~\ref{fig-alg-backward} details the backward search for a
pattern $\PATT$ using the condensed BWT {\cL} and corresponding
counts $\cC$.
As for regular backward search, an interval is maintained, initially
$\bsstate{\epsilon}{0}{0}{n-1}$.
That interval is then narrowed using the condensed arrays, adding one
more character into the matched string at each iteration of the loop.
The search commences with the rightmost symbol in the reverse of
$\PATT$, which is the leftmost symbol in $\PATT$; and (in the frame
of reference established in Figure~\ref{fig-rosa-backward}) prepends
subsequent matched characters to the left.
In particular, the search process maintains
\[
\lb = \min \left\{k \mid \TEXT[\SUF[k]..\SUF[k]+d-1]=\PATT[0..{d-1}] \right\}
\]
as the first suffix in $\SUF$ that matches $\PATT$ to depth $d$, and
\[
\rb = \max \left\{k \mid \TEXT[\SUF[k]..\SUF[k]+d-1]=\PATT[0..{d-1}] \right\}
\]
as the last such suffix.

To step from one configuration to the next, symbol $\PATT[d]$ must be
processed, with $\lb$ and $\rb$ updated so that the assignment
$d\leftarrow d+1$ then restores the invariant.
To narrow the $(\lb,\rb)$ interval the process described by Ferragina
and Manzini~\cite{fm05jacm} is used, but with an added level of
complexity: $\lb$ and $\rb$ are first translated into the condensed
domain, then processed against the condensed BWT $\cL$ in that
domain, and finally translated back to the full domain, ready for the
next iteration.
Those transformations are guided by two bitvectors $\bwdbl$ and
$\bwdbf$ (see the example in Figure~\ref{fig-rosa-backward}), which
record, respectively, which of the suffixes are needed during the
search in the condensed BWT, and where the lead symbols of those
suffixes appear in the full BWT string.
Rank and select operations on those two bitvectors yield the
operation sequence shown in Figure~\ref{fig-alg-backward}.

For example, to match $\PATT=\qstring{she}$, the first iteration
processes the {\qstring{s}}, and the configuration becomes
$\bsstate{\qstring{s}}{1}{12}{16}$.
Then a second iteration in which the {\qstring{h}} is processed
results in the configuration $\bsstate{\qstring{hs}}{2}{6}{7}$.
Now the interval is smaller than $b=3$, so the in-memory search is
ended, and the indicated suffix block (backward identifier $7$,
forward identifier $9$) is fetched.
A search for {\qstring{shy}} would also require that block $9$ be
accessed before the search could be declared a failure.
On the other hand, the pattern {\qstring{say}} generates the (condensed
domain equivalent of the) empty configuration
$\bsstate{\qstring{as}}{2}{3}{2}$ at step~05 after two iterations,
and reports failure at step~07 without a suffix block being required.

\begin{figure}[t]
\begin{tabbing}
\quad \=\quad \=\quad \=\quad \=\quad \=\quad \=\quad \=\quad \=\quad \=\quad \=\quad \=\quad \kill \\
\scriptsize 00 \>\> $\var{get\_bwd\_id}(\lb, d)$\\
\scriptsize 01 \>\>\> $\runnr \leftarrow \rank(\bwdbf, \lb, \qstring{1})$\\
\scriptsize 02 \>\>\> \Keyw{if} $\runnr = 0$ \Keyw{then}\\
\scriptsize 03 \>\>\>\> \Keyw{return} $0$\\
\scriptsize 04 \>\>\> $\runpos \leftarrow
		\select(\bwdbm, \runnr-1, \qstring{1})+1$\\
\scriptsize 05 \>\>\> $x \leftarrow \bwdmindepth[\rank%
		(\bwdbm, \runpos, \qstring{10})]$\\
\scriptsize 06 \>\> \> \Keyw{return} $\runpos - \runnr +%
		(d-x)$
\end{tabbing}
\mycaption{Determining the block identifier matching a
reverse search configuration $\bsstate{\omega}{d}{\lb}{\rb}$.
\label{fig-alg-backwardid}
}
\end{figure}

\begin{table*}[t]
\begin{center}
\mycaption{Structures required in memory during {\rosa} pattern
matching.
The value $z$ is the number of entries in each of $\bwdbf$ and
$\bwdbl$.
If there are $B$ suffix blocks, then $z\le\min\{4B, n\}$.
The final two columns show the actual cost for test file {\bigfile},
described in Table~\ref{tbl-datafiles}, and that number expressed as a
multiple of $B\log n$ bits, with $b=4{,}096$, and
$B=219{,}319{,}568$ blocks generated.
\label{tbl-memoryspace}}
\begin{tabular}{@{\extracolsep{4pt}}lllllcc}
\hline
Structure & Type & Operations & Parameters &
Space (upperbound, bits) & Space (actual, MB) & ${}\times B\log n$\\
\hline
$\bwdbf$ & bitvector & \select
	& $z$ elements, each $0\le x\le n$
	& $z(2+\log(n/z)) + \order{z}$
	& 135.3 & 0.144
\\
$\bwdbl$ & bitvector & \rank
	& $z$ elements, each $0\le x\le n$
	& $z(2+\log(n/z)) + \order{z}$
	& 135.3 & 0.144
\\
$\bwdbm$ & bitvector & \rank/\select
	& $2B$ elements, each $0\le x\le n$
	& $2B(1+\log(n/B)) + \order{B}$
	& \D37.4 & 0.040
\\
$\bwdmindepth$ & array & \access
	& $B$ elements, each $0\le x\le n-B$
	& $B\log n$
	& \D72.3 & 0.077
\\
$\cC$ & array & \access
	& $\sigma$ integers, each $0\le x<z$ 
	& $\sigma\log n$
	& \D\D$<$0.1\hphantom{$<$} & $<$0.001\hphantom{$<$}
\\
$\cL$ & array & \rank
	& $z$ symbols, each $0\le x<\sigma$
	& $\Order{z H_0(\cL)} = \Order{z\log\sigma}$
	& \D74.1 & 0.079
\\
\it pointers & array & \access
	& $B$ elements, each $0\le x \le n$
	& $B\log n$
	& 967.4 & 1.023
\\
\hline
\end{tabular}
\end{center}
\end{table*}

\begin{table*}[t!]
\mycaption{Details of data files.
The value of $H_k$ is empirical, generated by executing
{\tt{xz --best}}.
\label{tbl-datafiles}}
\begin{center}
\newcommand{\tabent}[1]{\makebox[15mm][c]{#1}}
\begin{tabular}{l cccc c ccc}
\hline
\multirow{2}{*}{Name}
	& \multirow{2}{*}{Type} & Size & 
	\multirow{2}{*}{$\sigma$} & $H_k$
	&& \multicolumn{3}{c}{LCP}
\\
\cline{7-9}
	& & (MB) && (bits/char) 
	&& \tabent{Median} & \tabent{Average} & \tabent{Maximum, $\hat{n}$}
\\
\hline
\tnyfile 
	& HTML/Web &\D\C256 & 129 & 0.45 %
	&&\C141 &\D\D5,937 &\D\C556,673
\\
\smlfile 
	& HTML/Web &\D4{,}000 & 129 & 0.57 %
	&&\C281 &\D11,506 &\D\C692,160
\\
\bigfile 
	& HTML/Web & 64{,}002 & 129 & 0.61%
	&& 1,896 &\D20,500 &\D1,204,953
\\
\dnafile 
	& Text/Genomic &\D2,985 &\D\D9 & 1.65 %
	&& \D\C16 & 554,171 & 29,999,999
\\
\dblpfile
	& XML/Bibliographic &\D1,032 &\D99 & 0.90 %
	&&\D\C36 &\D\D\C\D45 &\D\D\C\D1,353
\\
\hline
\end{tabular}

\end{center}
\end{table*}

\myparagraph{Computing Block Numbers}

Once a configuration $\bsstate{\omega}{d}{\lb}{\rb}$ has been
established by {\var{get\_interval}$()$}, the next step is to map it
to a block number; that is, identify the correct gray superscript
value associated with the black block identification circles in
Figures~\ref{fig-rosa-forward} and~\ref{fig-rosa-backward}.
Because multiple blocks might map to the same $\lb$ value but with
different depths $d$, a further bitvector {\bwdbm} is required,
containing a $0$-bit for each block in the forward suffix array, plus
a $1$-bit for each $1$-bit in {\bwdbf}, corresponding to blocks in
the reversed suffix array.
The bits are interleaved so that each entry point in {\bwdbm} is
preceded by a string of 0-bits that indicates the number of disk
blocks converging at that entry point.
The process of mapping via that structure, plus another array of
integers that records the minimum configuration depths at each valid
entry point, is described in Figure~\ref{fig-alg-backwardid}.

Once a block number in the reverse suffix has been identified, it is
converted to an on-disk byte address via an array storing a mapping
that is many-to-one because of the reducible blocks.
The configuration $\bsstate{\omega}{d}{\lb}{\rb}$ is then compared
with the $(\lb,d)$ values stored in the block's header, to identify
the matching $(\Delta_x,\Delta_d)$ region or subregion of the block
at which the search should be resumed.

\myparagraph{Space Requirement}

The bitvectors and arrays required in memory during querying are
summarized in Table~\ref{tbl-memoryspace}.
The symbols extracted into the condensed BWT are exactly those
required during searching for any of the block prefix strings.
No BWT symbols that would only be accessed if $\rb-\lb$ was permitted
to become smaller than $b$ are needed.
At most two bits are required for each node in the corresponding
frequency-pruned suffix tree, and that tree contains at most $2B$
nodes if the {\rosa} contains $B$ disk blocks.
The maximum number of bits that can be set is $n$, meaning that the
actual number of bits set, $z$, is bounded by $z\leq\min\{4B,n\}$.
When $b$ is large, $B$ can be expected (but not guaranteed) to be
small, making the bitvectors $\bwdbf$ and $\bwdbl$ sparse and highly
compressible; and making the $\cL$ and $\cC$ arrays that represent
the condensed BWT small too.

\myparagraph{Execution Time}

Function~$\var{get\_interval}()$ in Figure~\ref{fig-alg-backward}
iterates at most once for each character in the pattern.
A total of two bitvector {\rank} operations and two bitvector
{\select} operations are required per iteration; each of these take
$\Order{1}$ time.
Step~05 involves {\rank} operations on an array, $\cL$.
That array is implemented as a Huffman-shaped wavelet tree, based on
underlying bitvectors, meaning that symbol-based {\rank} queries can
be carried out via not more than $\log\sigma$ bitvector-based {\rank}
queries, or in $\Order{\log\sigma}$ time .
The process of finding the matching block identifier (function
$\var{get\_bwd\_id}()$ in Figure~\ref{fig-alg-backwardid}) involves
only rank and select operations on bitvectors, and takes $\Order{1}$
time per pattern.

We now bring together these various observations, and state the main
result of this section.

\bigskip\noindent
{\sc{Theorem 1}}:
{\emph{Given a set of $B$ strings corresponding to the leaves of a
pruned suffix tree for a text of $n$ symbols, the condensed BWT
structure
requires $\Order{(B+\sigma)\log
n}$ bits of storage and
identifies the leaf corresponding to an $m$-symbol pattern
in $\Order{m\log\sigma}$ time.
}}

\section{Experiments}
\label{sec-experiments}

We have implemented and tested our Reduced On-disk Suffix Array, or
{\rosa}, and compared it against a range of alternatives.

\myparagraph{Experimental Hardware}

Experiments were run on two different hardware
platforms:
a MacBook Pro with a $2.4$~GHz Intel Core i5 processor, {\gb{4}} RAM,
and {\gb{500}} hard disk; and a MacBook Air with $1.8$~GHz Intel Core
i7 processor, {\gb{4}} RAM, and a {\gb{250}} solid-state disk.
The suffix array itself was prepared on a separate server with a
large amount of main memory.

\myparagraph{Test Data}

Data was obtained from a range of sources, with an emphasis on large
files.
The first suite of test files were drawn from the 2009
{\method{ClueWeb}} collection, a large-scale web
crawl\footnote{\url{http://lemurproject.org/clueweb09.php/}}.
Three files were extracted as prefixes of the concatenation of the
first $64$ files in the directory
{\url{ClueWeb09/disk1/ClueWeb09_English_1/enwp00/}}, with null bytes
in the text replaced by 0xFF-bytes.
(Null byte is the {\qstring{\$}} symbol reserved in all our
implementations to mark the end of the input string.)
In Table~\ref{tbl-datafiles} these three files are denoted as
{\tnyfile}, {\smlfile}, and {\bigfile}.
Two other types of data were also used: file {\dnafile} is a text
file representing the human genome stored as a sequence of ASCII
letters (primarily {\qstring{A}}, {\qstring{C}}, {\qstring{G}}, and
{\qstring{T}}); and file {\dblpfile} is an XML repository containing
$844{,}702$ bibliographic references to computing research
papers\footnote{\url{http://dblp.uni-trier.de/xml/}}.

The three different types of data differ markedly in the extent to
which they contain sequence repetitions.
In the web data the LCP values are particularly high, caused by reuse
of formatting text, and by duplicate documents.
The median LCP is much lower for the XML and DNA data; but note that
the file {\dnafile} contained a repeated subsequence of thirty
million characters.
The three data types also differ in the size of the alphabet used,
and in compressibility.
To estimate the latter quantity, the column marked $H_k$ shows the
compression achieved by a high-quality mechanism, expressed in terms
of bits per character relative to the original.
The web and XML data are highly compressible; the DNA file somewhat
less so.

\myparagraph{Test patterns}

To generate test queries, a suffix tree representation of each file
was processed sequentially, and a large set of $\langle$pattern,
frequency$\rangle$ pairs identified.
These were then quantized by both pattern length and by pattern
frequency, with agreement assumed in the second dimension if the
actual frequency was within $25$\% of one of a set of target
frequencies.
This approach allowed a total of $25$ different query sets to be
formed for each file, representing all combinations of
$|\PATT|\in\{4,10,20,40,100\}$ and pattern frequency $k\in\{10^0,
10^1, 10^2, 10^3, 10^4\}$.
On the web data, all combinations occurred more than $1{,}000$ times,
and experiments were run on random subsets of size $1{,}000$ drawn
from the corresponding category.
Selected combinations of $|\PATT|$ and $k$ were used for the other
datafiles, and results are similarly the average over $1{,}000$
patterns.
It was not possible to identify any patterns with $|\PATT|=4$ and
$k=10{,}000$ on {\dnafile}, and as a result one entry is omitted in
the tables below.

\begin{figure}[t]
\begin{center}
\begin{tabular}{ccc}
\includegraphics[scale=1.05]{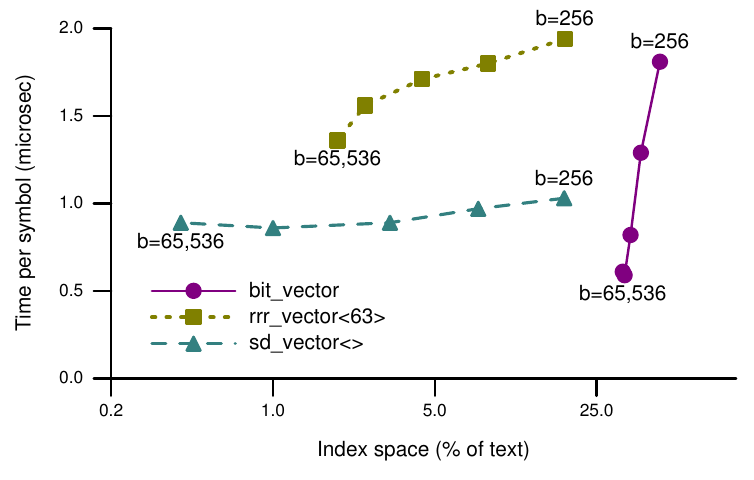}
\end{tabular}
\end{center}
\mycaption{Space and processing time of in-memory search using
condensed BWT approach as a function of blocksize, for three
different bitvector representations.
Data is for {\tnyfile}, averaged over $1{,}000$ patterns with
$|\PATT|=40$ and $k=10{,}000$ matches per pattern, and with the
blocksize varying between $b=2^8$ and $b=2^{16}$.
\label{fig-inmemorypart}}
\end{figure}

\myparagraph{Compressed Bit Vectors}

\begin{table}
\mycaption{Percentage of index space required by components of
condensed BWT index for {\smlfile}.
\label{tbl-condensed}}
\begin{center}
\begin{tabular}{l c ccc}
\hline
Component 
	&& $b=2^8$ & $b=2^{12}$ & $b=2^{16}$
\\
\hline
Bitvectors ({\bwdbf}, {\bwdbl}; SD-array)
	&& 14.3 & 22.4 & 31.7
\\
Condensed BWT ({\cL}; wavelet tree)
	&&\D5.2 &\D6.5 &\D7.8
\\
Auxiliary information
	&&\D7.7 &\D8.9 &\D9.7
\\
Pointers (binary)
	&& 72.8 & 62.2 & 50.7
\\
\hline
\end{tabular}
\end{center}
\end{table}

\begin{table*}
\mycaption{In-memory search structures for variable suffix array
blocks.
\label{tbl-inmemory}}
\begin{center}
\begin{tabular}{lc c cc c cc}
\hline
\multirow{2}{*}{Data}
	& \multirow{2}{*}{$b$}
	&& \multicolumn{2}{c}{Memory (MB)}
	&& \multicolumn{2}{c}{Query speed (microseconds/query)}
\\
\cline{4-5}\cline{7-8}
	&
	&& Condensed BWT & Bit-blind tree
	&& Condensed BWT & Bit-blind tree
\\
\hline
\smlfile
	& $2^{10}$
	&& 269.8 & 329.1 
	&& 33.7  & 36.3
\\
\smlfile
	& $2^{12}$
	&&\D98.6 & 112.2
	&& 24.3  & 31.5
\\
\smlfile
	& $2^{14}$
	&&\D15.8 &\D15.1
	&& 19.7  & 28.6
\\[1ex]
\dblpfile
	& $2^{10}$
	&&\D58.3 &\D58.4
	&& 26.8  & 29.4
\\
\dblpfile
	& $2^{12}$
	&&\D21.1 &\D18.9
	&& 19.2  & 24.2
\\
\dblpfile
	& $2^{14}$
	&&\D\D7.6 &\D\D6.2
	&& 15.6  & 19.9
\\[1ex]
\dnafile
	& $2^{10}$
	&& 410.2 & 382.8
	&& 29.7  & 24.3
\\
\dnafile
	& $2^{12}$
	&& 342.9 & 319.3
	&& 21.1  & 21.0
\\
\dnafile
	& $2^{14}$
	&& 326.6 & 307.8
	&& 17.8  & 17.3
\\
\hline
\end{tabular}
\end{center}
\end{table*}

A key decision is how to represent the two large bitvectors.
Conceptually each of them contains $n$ bits, but, by construction,
the number of $1$ bits is close to the number of suffix array disk
blocks, and so they are sparse and amenable to compression.
The drawback of compression is that {\rank} and {\select} operations
become slower.
Figure~\ref{fig-inmemorypart} compares the space and access cost of
three different representations for the two bitvectors, with space
plotted on the horizontal axis, measured as the ratio of the complete
condensed BWT data structure as a fraction of the text size; and
processing time per matched character plotted vertically.
The alternatives are denoted by their {\tt{sdsl}} class
identifiers\footnote{{\url{https://github.com/simongog/sdsl}}}:
uncompressed bitvectors (class {\sdslbitvector}); the well-known RRR
structure~\cite{rrr02soda} ({\sdslrrrvectorvar}); and the SD-array
({\sdslsdvector}) of Okanohara and Sadakane~\cite{os07alenex}.
The SD-array offers the best balance, and while it is not always
faster than the uncompressed bitvector alternative, it occupies much
less space.

Once the bitvectors are compressed, the disk block pointers are the
most costly component of the condensed BWT index.
These are addresses into the index (for irreducible and reducible
blocks) or into the text (for singletons), and are represented as
minimal-width binary numbers.
Table~\ref{tbl-condensed} shows the percentage of the total memory
space required by each of the four main components of the condensed
BWT search structure, for the file {\smlfile} and three different
blocksizes.
The dominance of the pointers is clear.

\myparagraph{Baseline Methods and Total Disk Space}

In any experimental comparison it is important to compare against
appropriate reference points.
The {\rosa} structure -- consisting of condensed in-memory BWT array
index, and a reduced set of suffix array blocks stored on disk, can
be compared with the {\lofsa} (which in turn is compared by Sinha et
al.~\cite{spmt08sigmod} against previous data structures); with the
{\sbtree}; and with the {\fmindex}.
The {\fmindex} is not a two-level disk-based mechanism, and can only
be used if the complete structure fits main memory.
Nevertheless, it is substantially smaller than the other structures,
meaning that its zone of applicability is larger, and overlapping with
the size range for which two-level structures are appropriate.

Table~\ref{tbl-compare-sizes} compares index sizes for these various
approaches, including both components for the two-level ones.
The values for the {\rosa} and {\fmindex} are measured based on our
experimental implementations.
There is no software for the {\sbtree} or {\lofsa} capable of
handling the data sizes used in our experiments, and the values shown
in the table marked with ``*'' are computed using
Equation~\ref{eqn-sbtree} (in Section~\ref{sec-ondisksuffix}) for the
{\sbtree}, and estimated from the results given by Sinha et
al.~\cite{spmt08sigmod,mps09dasfaa} for the {\lofsa}.
With the exception of the {\fmindex}, all of these structures require
that the text $\TEXT$ also be stored, adding a further {\gb{3.9}}.

As can be seen, the block reductions achieved in the {\rosa} mean
that it is by far and away the smallest of the two-level approaches.
Indeed, the {\rosa} index requires just half the space of a plain
suffix array.
On the other hand, the {\sbtree} and the {\lofsa} are expensive to
store; neither of these structures support block reductions, and in
the case of the {\sbtree}, the LCP values are also a costly component
because the fixed block structure means that they cannot be stored
compressed.
Because of their clear space superiority, the remainder
of the experimentation focuses on the {\rosa} and the {\fmindex}
alone.

\begin{table}[t]
\mycaption{Total memory and disk space required for two-level suffix
array structures and the {\fmindex}, for {\smlfile}.
\label{tbl-compare-sizes}}
\begin{center}
\begin{tabular}{lc l c c}
\hline
Structure && Ref. && Size (GB)\\
\hline
Suffix array &
	& \cite{mm93siamjc}   && 15.6\hphantom{*}\\
{\lofsa} &  $b=4{,}096$
	& \cite{spmt08sigmod} && 46.9*\\
{\lofsa} &  $b=4{,}096$
	& \cite{mps09dasfaa}  && 27.3*\\
{\sbtree} &  $b=4{,}096$
	& \cite{fg99jacm}     && 24.5*\\
{\rosa} &  $b=4{,}096$
	& \it this paper        && \D7.8\hphantom{*}\\
{\fmindex} &
	& \cite{fm05jacm}     && \D0.6\hphantom{*}\\
\hline
\end{tabular}
\end{center}
\end{table}

\myparagraph{Choice of In-Memory Structure}

The second step of the experimental evaluation was to compare the
condensed BWT method with the bit-blind tree, in terms of memory
space required and search time to identify suffix blocks
(Table~\ref{tbl-inmemory}).
Search times are measured over frequently-occurring long queries
($|\PATT|=40$ and $k=10{,}000$) (so that the search is driven towards
the extremities of the in-memory structure); and include only the
cost of processing the in-memory data structure.

The two methods are comparable in their space requirements, with the
bit-blind tree sometimes being a little smaller, and the condensed
BWT structure sometimes being a little smaller.
The condensed BWT has a small but consistent advantage in terms of
CPU time.
Search in the condensed BWT structure requires fewer loop iterations
than in the bit-blind tree, but each iteration is more expensive.
Note that Table~\ref{tbl-inmemory} does not include the cost of the
disk accesses to $\TEXT$ needed to resolve the uncertainty inherent
in the bit-blind search process.
Details of disk access costs are presented shortly; the condensed BWT
arrangement has a clear advantage when that cost is included.

\myparagraph{Blocksizes and Non-Uniform Sampling}

\begin{figure}[t]
\begin{center}
\includegraphics[scale=1.05]{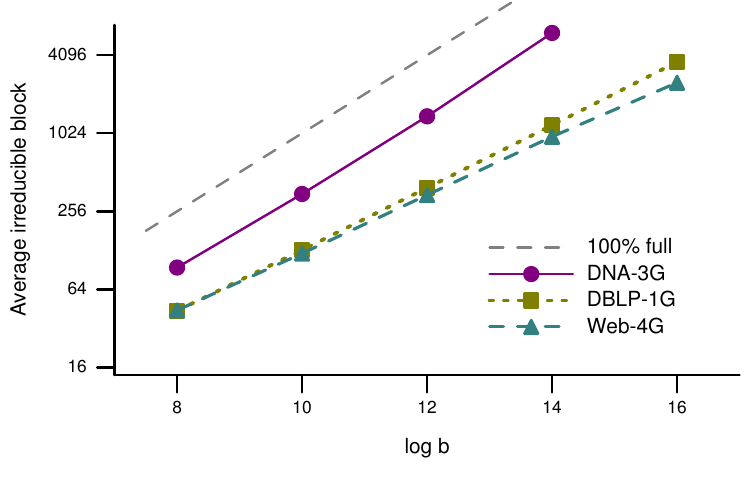}
\mycaption{Average size of irreducible blocks (in pointers).
\label{fig-block-sizes}}
\end{center}
\end{figure}

Figure~\ref{fig-block-sizes} depicts the average number of pointers
stored in each irreducible block for three of the test files.
The growth in average block size is linear in the size of the block,
but for the non-genomic data the average is well below the
limit $b$.
This relationship is not unexpected -- blocks are formed at nodes of
the suffix tree whenever the parent has a count of more than $b$, but
the node in question does not.
At that boundary node, the available symbol count is split across all
of the children.
When the alphabet size $\sigma$ is large, those child counts will, on
average, be relatively small.
The same observation explains why the blocks are larger for the file
{\dnafile} -- when $\sigma$ is small, the average frequency count in
each child is likely to be larger.

\begin{figure}[t]
\begin{center}
\includegraphics[scale=1.05]{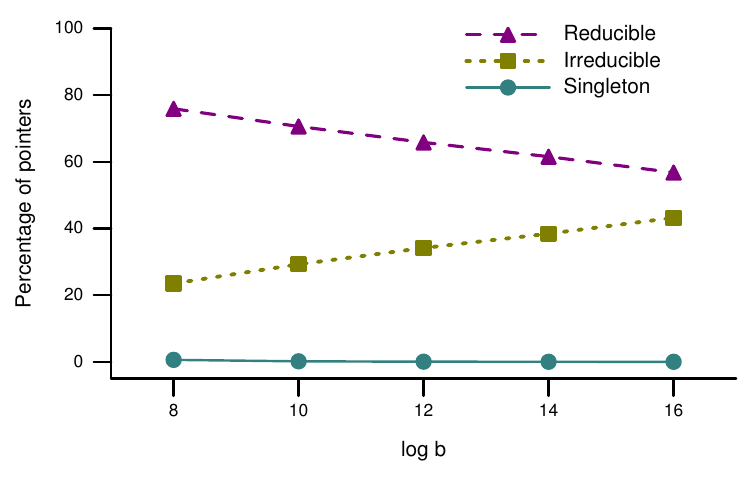}
\mycaption{Fraction of pointers in reducible, irreducible, and
singleton blocks for
{\smlfile}
and different values of $b$.
\label{fig-pointers}}
\end{center}
\end{figure}

Figure~\ref{fig-pointers} shows the fraction of the suffix pointers
located in reducible blocks, irreducible blocks, and
singleton blocks for {\smlfile}.
When $b$ is small, more than two thirds of the suffix pointers are in
reducible blocks.
That fraction decreases as $b$ increases, not because the
reductions are no longer present, but because the similar sections
no longer span whole blocks.
But even when $b=65{,}536$, around half of the suffix pointers can be
eliminated.
Similar behavior was observed for {\dblpfile}.
On the other hand, the DNA data has markedly different
characteristics, and while it generates many more singleton blocks,
the number of block reductions is very small.

\begin{table}[t]
\mycaption{Space required by {\rosa} query-time index components
with $b=4{,}096$,
expressed as multiples of the source text size.
\label{tbl-space}}
\begin{center}
\newcommand{\tabent}[1]{\makebox[12mm][c]{#1}}
\begin{tabular}{l c cc}
\hline
File & \tabent{Memory} & \tabent{Disk} & \tabent{Total, inc.~$\TEXT$}\\
\hline
\tnyfile
	& 0.033 & 1.943 & 2.976
\\
\smlfile
	& 0.025 & 1.961 & 2.986
\\
\bigfile
	& 0.022 & 1.900 & 2.922
\\[1ex]
\dblpfile
	& 0.020 & 2.126 & 3.146
\\[1ex]
\dnafile
	& 0.116 & 4.704 & 5.820
\\
\hline
\end{tabular}
\end{center}
\end{table}

Table~\ref{tbl-space} shows the balance between in-memory space and
on-disk space required by the {\rosa} for the full set of data files.
For the web and XML data, the total space required is much less than
would be required by a plain suffix array (which is a factor of
$4.75$ for {\dblpfile}, and of $5.0$ for {\smlfile}).
On the other hand, the {\rosa} handles the DNA data relatively
poorly, and both the in-memory index and the on-disk component are
large.
Indeed, on the DNA data the {\rosa} takes more space than a plain
suffix array, a consequence of the relative absence of
repetitions.

\begin{table}[t]
\mycaption{Disk accesses per count query for file {\smlfile},
with $b=4{,}096$.
\label{tbl-disk}}
\begin{center}
\newcommand{\tabent}[1]{\makebox[8mm][c]{#1}}
\begin{tabular}{c}
\begin{tabular}{l ccccc}
\hline
\multirow{2}{*}{$|\PATT|$}  & \multicolumn{5}{c}{Number of answers}\\
\cline{2-6}
	&    
	\tabent{1} &  
	\tabent{10} & 
	\tabent{100} &
	\tabent{1,000} &
	\tabent{10,000}\\
\hline
\D\D4  & 1.79 & 1.52 & 1.12 & 0.35 & 0.00\\ 
\D10   & 1.99 & 1.99 & 1.94 & 1.70 & 0.00\\ 
\D20   & 2.00 & 1.99 & 1.98 & 1.83 & 0.00\\ 
\D40   & 2.00 & 2.00 & 1.99 & 1.90 & 0.00\\ 
 100   & 2.00 & 2.00 & 2.00 & 1.95 & 0.00\\ 
\hline
\end{tabular}
\scriptsize (a) Condensed BWT\\[1ex]
\begin{tabular}{l ccccc}
\hline
\multirow{2}{*}{$|\PATT|$}  & \multicolumn{5}{c}{Number of answers}\\
\cline{2-6}
	&    
	\tabent{1} &  
	\tabent{10} & 
	\tabent{100} &
	\tabent{1,000} &
	\tabent{10,000}\\
\hline
\D\D4  & 1.86 & 2.00 & 2.00 & 2.00 & 1.84 \\ 
\D10   & 1.99 & 2.00 & 2.00 & 2.00 & 1.87 \\ 
\D20   & 2.00 & 2.00 & 2.00 & 2.00 & 1.90 \\ 
\D40   & 2.00 & 2.00 & 2.00 & 2.00 & 1.87 \\ 
 100   & 2.00 & 2.00 & 2.00 & 2.00 & 1.94 \\ 
\hline
\end{tabular}
\scriptsize (b) Bit-blind tree
\end{tabular}
\end{center}
\end{table}

\begin{table*}[t]
\newcommand{\tabent}[1]{\makebox[14mm][c]{#1}}
\mycaption{Execution times in milliseconds per query, using two different
hardware platforms, with $b=4{,}096$.
\label{tbl-querytime}}
\begin{center}
\begin{tabular}{l l ccccc}
\hline
\multirow{2}{*}{Text} &
	\multirow{2}{*}{Platform} &
	$|\PATT|=4$ &
	$|\PATT|=10$ &
	$|\PATT|=20$ &
	$|\PATT|=40$ &
	$|\PATT|=100$
\\
	&&
	\tabent{$k=10{,}000$} &
	\tabent{$k=1{,}000$} &
	\tabent{$k=100$} &
	\tabent{$k=10$} &
	\tabent{$k=1$}
\\
\hline
\multicolumn{7}{l}{\emph{Using the {\rosa}}}\\
{\dblpfile} & MacBook Air, SSD
	&\D\D\D0.004 &\D\D\D1.02 &\D\D\D1.10 &\D\D\D1.09 &\D\D\D1.13
\\
{\dnafile} & MacBook Air, SSD
	&\D\D---   &\D\D\D0.72 &\D\D\D1.10 &\D\D\D1.15 &\D\D\D1.23
\\
{\smlfile} & MacBook Air, SSD
	&\D\D\D0.006&\D\D\D1.00&\D\D\D1.06 &\D\D\D1.06 &\D\D\D1.05 
\\
{\bigfile} & MacBook Air, SSD
	&\D\D\D0.009&\D\D\D0.98&\D\D\D1.04&\D\D\D1.09 &\D\D\D1.13
\\[1ex]
{\dblpfile} & MacBook Pro, mechanical disk
	&\D\D\D0.005&\D\D21.1\D&\D\D25.5\D&\D\D24.8\D&\D\D26.5\D
\\
{\dnafile} & MacBook Pro, mechanical disk
	&\D\D---   &\D\D14.9\D&\D\D25.3\D&\D\D25.8\D&\D\D26.7\D
\\
{\bigfile} & MacBook Pro, mechanical disk
	&\D\D\D0.009 &\D\D33.9\D&\D\D40.3\D&\D\D40.7\D&\D\D44.6\D
\\[1ex]
\multicolumn{7}{l}{\emph{Using an efficient {\fmindex}}}\\
{\smlfile} & MacBook Air, SSD
	&\D\D\D0.011&\D\D\D0.03&\D\D\D0.07&\D\D\D0.14&\D\D\D0.36   %
\\	
{\bigfile} & MacBook Air, SSD
	&\D\D44.6\D\D &\D\D85.6\D &\D\D88.9\D &\D118.9\D &\D\D70.0\D
\\
{\bigfile} & MacBook Pro, mechanical disk
	&\D630\hphantom{.00}&1450\hphantom{.00} & 2040\hphantom{.00} & 2500\hphantom{.00} & \D980\hphantom{.00}
\\
\hline
\end{tabular}
\end{center}
\end{table*}

\myparagraph{Disk Accesses and Execution Cost For Count Queries}

Table~\ref{tbl-disk} shows the number of disk accesses required by
the two options for the in-memory structure.
The benefit of the condensed BWT arrangement is clear -- because it
admits no ambiguity, fewer disk accesses are required for {\qcount}
queries when the pattern is common in the text and can be resolved
entirely within the in-memory index.
When the pattern is frequent, the discrepancy is even greater -- the
condensed BWT allows {\qcount} queries to be processed without
recourse to disk, whereas the bit-blind tree still requires an average of
more than $1.8$ disk accesses per query.

Table~\ref{tbl-querytime} shows overall elapsed times for a range of
query lengths and frequencies across the set of data files (including
the {\gb{64}} file), and for two hardware platforms.
The in-memory condensed BWT index for {\bigfile} requires {\gb{1.39}}
(around two-thirds of which is pointers, as shown in the final column of
Table~\ref{tbl-memoryspace}),
and the on-disk part a total of {\gb{119}}, with the latter composed
of {\gb{1.4}} for block headers and other auxiliary data; {\gb{29.5}}
for compressed LCP differentials and for tree structure bits; and
{\gb{82.7}} for suffix pointers.
Including the text $\TEXT$, the entire search system requires
{\gb{183}}, a factor of $2.9$ relative to the text, and only a little
over half of the $5.5$-factor that would be required by a simple
suffix array, not even including any allowance for LCP values.

As can be seen, access via SSD memory is much faster than access via
mechanical disk.
But even with the mechanical disk, pattern queries on {\bigfile} can
be answered by the {\rosa} in under 50~milliseconds.
Moreover, search times are largely unaffected by pattern length,
except that queries on frequently-occurring strings are always
handled within a small number of microseconds.

\myparagraph{Compared to the {\fmindex}}

The last three rows of Table~\ref{tbl-querytime} show the query cost
of a highly-tuned (for both space and speed) {\fmindex}
implementation that has been demonstrated to outperform other
available packages {\cite[Section 6.6]{gp13spe}}.
For {\smlfile}, a run-length compressed wavelet tree and SD-array
implementations for the two {\fmindex} bitvectors was used, the
fastest configuration.
During querying, this {\fmindex} version requires {\mb{659.4}} of
memory space.
For short {\qcount} queries it is much faster than the {\rosa}.
With a different bitvector representation (using the RRR variant),
space can be reduced to {\mb{404.6}}, but querying time increases by
a factor of around three.

For {\bigfile} (the last two lines of Table~\ref{tbl-querytime}), the
more compact RRR bitvector option was used, requiring {\gb{8.3}} for
the index.
As can be seen, when only a subset of a large index can be maintained
permanently in memory, the non-sequential access pattern means that
retrieval times increase dramatically.
When SSD disk is used the times are still somewhat plausible, but the
two-second response times that arise when a mechanical disk is used
are anything but plausible.
The sequence of results in Table~\ref{tbl-querytime} clearly
highlights the situations for which the {\rosa} is the fastest search
mechanism.

\section{Discussion}
\label{sec-discussion}

We conclude by comparing the {\rosa} with other large-scale
search mechanisms that have been presented in the literature.

\myparagraph{Construction and Applicability}

Despite recently developed techniques~\cite{bfo13alenex}, a drawback
of all suffix array-based pattern search methods is the cost of
building the suffix array.
The structures used in our experiments were generated on a server
with considerably more memory than the laptops that were used for the
search experiments, and reflect the situation for which we believe
static two-level structures are best suited -- namely, when large
fixed texts are to be pre-processed by a central service to make
``searchable packages'' that can be distributed onto low-cost devices
for querying purposes.

The {\fmindex} is a strong competitor for the same type of
applications.
It has approximately the same construction cost, but a much smaller
query-time disk storage footprint.
The disadvantage of using an {\fmindex} is that for any given text
$\TEXT$, its memory requirement is likely to be greater than that of
the {\rosa}, because the entire structure must be present in memory.
That is, there is a size of text for which an {\fmindex} cannot be
supported by the available hardware, but a {\rosa} can, albeit with
significantly greater disk storage consumption.
Depending on the exact configuration used, {\qlocate} and {\qcontext}
queries might also be slower in an {\fmindex} than using the {\rosa}.

It is also interesting to calculate the break-even point at which a
pre-computed data structure becomes more economical than sequential
search.
Construction of the {\rosa} for {\smlfile} requires around $100$
minutes, and the current implementation involves a peak memory
requirement of $9n$ bytes during the two suffix sorting steps
(external methods for suffix sorting are available that reduce the
memory cost, but increase the construction time).
Using the MacBook Pro to search the same {\gb{4}} file for patterns
using {\tt{agrep}}\footnote{\url{ftp://ftp.cs.arizona.edu/agrep/}.}
requires about
three seconds, once the file containing $\TEXT$ has been brought in
to memory.
Hence, construction of a {\rosa} index is warranted if more than
around $2{,}000$ queries are to be processed against the same text $\TEXT$.

\myparagraph{Other Recent Work}

Phoophakdee and Zaki~\cite{pz07sigmod} describe a partition/merge
approach to suffix tree construction that allows them to undertake
pattern search on a human genome.
They compare their {\method{Trellis}} approach to other options on
files of up to three billion DNA base pairs, with a build time of
under six hours, and a final size of {\gb{71.6}}, or $27$ times
larger than the input text.
Using their suffix tree, they are able to undertake queries of $100+$
base pairs in approximately $60$~milliseconds.

Wong et al.~\cite{wsw07icde} describe a partitioned suffix tree they
call a {\method{CPS-Tree}}.
They experiment with files of $118$ million base pairs and $4.6$
million base pairs, and obtain suffix trees that require between $7n$
and $9n$ bytes.
With these small test files, querying is fast -- of the order of
$20$ microseconds per query -- because it still takes place in main memory.

Orlandi and Venturini~\cite{ov11pods} have also described a structure
for storing a pruned suffix tree.
Their pruning definition differs from the one used in the {\rosa},
and they retain a node if its size is greater than $b$, whereas in
the {\rosa} a node appears in the condensed BWT structure if its
{\emph{parent}} is of size greater than $b$.
The difference means that care must be taken when comparing sizes for
a given parameter value, since the {\rosa} retains as many as
$\sigma$ times more tree nodes than does the {\sc{CPST}}, including,
for example, singleton blocks.

For a {\sc{CPST}} over $n$ symbols in which there are $K$ suffix tree
nodes retained each of size $b$ or more, the space required by
Orlandi and Venturini's structure is $\Order{K\log(\sigma b)
+\sigma\log n}$ bits.
Direct comparison with the costs shown in Table~\ref{tbl-memoryspace}
is not possible, because for any given value of $b$ the number of
nodes $K$ in the {\sc{CPST}} is much less than the number of leaves
$B$ in the {\rosa} index structure.
The {\rosa}'s condensed BWT index provides greater functionality,
since it retains frequency counts for $(\lb,\rb)$ intervals narrower
than $b$, whereas the {\sc{CPST}} replies to {\qlocate} and {\qcount}
queries on rare and non-existent patterns with a uniform answer of
``don't know, if $\PATT$ does exits, it appears fewer than $b$
times''.
The {\rosa} also stores disk block pointers, a component that is not
required in the {\sc{CPST}}.
Orlandi and Venturini~\cite{ov11pods} also describe a
uniform-sampling index in order to undertake approximate {\qcount}
queries, where the returned pattern frequency in {\qcount} queries is
correct to within an additive fidelity constraint determined at the
time the index is constructed.
Building a {\sc{CPST}} requires initial construction of a
suffix tree, and needs more resources than creation of the
BWT string, the basis of the {\rosa}'s construction process.

Other recent work is by Ferguson~\cite{ferguson12cpm}, who describes
a search structure called {\femto}, and provides experiments on
{\gb{43}} of English text (Project Gutenberg files), and on
{\gb{182}} of genomic data.
The {\femto} system uses a partitioned {\fmindex}, with the search
for each pattern proceeding through (at least) one disk block per
symbol.
Ferguson gives experimental results showing that the constructed
index requires as little as half of the space of the original file,
but with query response times of $1$--$3$ seconds for {\qcount}
queries against selected patterns of $12$-$28$ symbols (two to three
word phrases, with tests carried out on an individual basis on
hand-selected strings, rather than as part of a regime of extensive
measurement) against the English text when using a conventional disk
drive; and of $10$ or more seconds when searching the Genomic data for
patterns of length $128$.
The high search times arise because of the disk accesses.
When multiple queries are simultaneously active, and duplicate
requests for disk blocks can be batched and processed all at once,
throughput improves dramatically, but with a corresponding increase
in individual response times.
Compared to the {\femto}, the methods presented here require more
disk space for the suffix array data, but operate an order of
magnitude more quickly.

Another approach to large-scale pattern search is to index
overlapping $t$-grams from $\TEXT$, each containing $t$ consecutive
symbols.
In total, $n-t+1$ locations in $\TEXT$ are indexed via a vocabulary
containing at most $\Order{\sigma^t}$ entries.
An inverted index is built, storing a variable-length postings list
for each unique $t$-gram, and recording the locations in $\TEXT$ at
which that particular combination of $t$ symbols
appears~\cite{zm06compsurv}.
Queries of length $m>t$ are resolved by intersecting the relevant
postings lists, identifying locations at which fragments overlap in
the desired manner; queries of length $m\le t$ are resolved by taking
the union of the postings lists of the vocabulary entries that
contain $\PATT$ within the $t$-symbol identifier.

Inverted indexes allow queries to be resolved in two disk accesses
per query term, one to retrieve a block of the vocabulary, and one to
retrieve a postings list~\cite{wmb99mg}.
If $t$ is chosen so that the $t$-gram vocabulary for $\TEXT$
can be held in main memory, the number of disk accesses required to
match a pattern $\PATT$ and resolve {\qlocate} queries is $\lceil
m/t\rceil$.
In terms of space,
a $t$-gram index with $t \approx 5$ to $10$ can be expected to
consume around $150$--$200$\% of the space required by $\TEXT$, and
to grow larger as $t$ increases.
Note that in the $t$-gram approach to pattern search $\TEXT$ is not
required in memory.

Tang et al.~\cite{tsb09cikm} give details of the construction and use
of $n$-gram indexes for pattern matching.
Puglisi et al.~\cite{pst06spire} have also examined this problem.

\section{Summary}

We have carried out a detailed investigation of two-level
suffix-array based pattern search mechanisms, and: (1) described an
efficient mechanism for exploiting whole block reductions, to
approximately half the space required by the suffix array pointers;
(2) described and analyzed a condensed BWT mechanism for storing and
searching the string labels of a pruned suffix tree; and (3)
described a comprehensive approach to testing pattern search
mechanisms.
We have demonstrated that in combination the new techniques provide
efficient large-scale pattern search, requiring around half the disk
space of previous two-level techniques, and providing faster search
than an {\fmindex} when the data is such that the {\fmindex} cannot
be accommodated in main memory.
While we have focused on the memory-disk interface, we note
that structures with the properties exhibited by the {\rosa} are
effective across all interface levels in the memory hierarchy.

\section*{Acknowledgment}

This work was funded by the Australian Research Council.
The {\rosa} software will be made publicly available.

\ifCLASSOPTIONcaptionsoff
  \newpage
\fi

\bibliographystyle{IEEEtran}
\bibliography{local}

\end{document}